\RequirePackage{etex}
\documentclass{aa}  

\usepackage[utf8]{inputenc}
\usepackage{graphicx}
\usepackage{subcaption}
\usepackage{physics}
\usepackage{xcolor}
\usepackage{float}
\usepackage[normalem]{ulem}
\usepackage[switch]{lineno}
\usepackage{amsmath}
\usepackage{comment}
\usepackage{placeins}
\usepackage{txfonts}
\usepackage[hidelinks]{hyperref}
\usepackage{cleveref}
\usepackage{newunicodechar}
\newunicodechar{α}{\ensuremath{\alpha}}
\crefname{figure}{Fig.}{Figs.}

\newcommand{\ha}{{H$\alpha$ }}

\begin{document}

\title{FILOS: a solar Filament Oscillation Simulator}
   \subtitle{}

\author{G. Castelló
      \inst{1,2}
      \and
      M. Luna 
      \inst{1,2}
      \and
      J. Terradas
      \inst{1,2}
      }

\institute{Departament de Física, Universitat de les Illes Balears (UIB), E-07122 Palma de Mallorca, Spain\\
          \email{guillem.castello@uib.cat}
     \and
         Institute of Applied Computing \& Community Code (IAC3), UIB, E-07122 Palma de Mallorca, Spain}

   \date{Received October 10, 2026; accepted November 16, 2026}

\abstract
{Solar filament oscillations provide diagnostics of prominence structure and
stability, but methods for recovering their physical properties require
validation against observations for which the filament geometry and dynamics
are known.}
{We present FILOS, a solar Filament Oscillation Simulator that generates
synthetic filaments with prescribed oscillatory motions in H$\alpha$
observations representative of the Global Oscillation Network Group (GONG).}
{Filament morphology is constructed from measured spines and populated with
three-dimensional dipped threads. Each thread follows a circular magnetic dip
and contains hydrostatic prominence plasma, from which the H$\alpha$ optical
depth is computed. The filament is projected onto the image plane, degraded
according to the GONG instrumental response, and superimposed on observed
GONG H$\alpha$ background sequences. Longitudinal and transverse damped
motions can be imposed independently. Longitudinal periods may be prescribed
or computed for each thread from the dip curvature and prominence height using
the pendulum model.}
{FILOS produces GONG-like H$\alpha$ sequences with filament geometry and
dynamics known by construction. In a representative longitudinal experiment,
thread periods span 44.2--87.6 min (median 60.8 min). Standard-normalized Lomb–Scargle power maps reveal the dominant period and the expected period-halving signature near equilibrium.
A time--distance fit recovers an onset of $19.53\pm0.23$ min for a prescribed
20 min and a collective period of $57.71\pm0.35$ min. Differences between the fitted collective quantities and the nominal thread-level inputs are consistent with the effects of projection, image formation, and progressive thread dephasing. FILOS also provides segmentation masks and opacity-weighted plane-of-sky velocity
fields.}
{FILOS provides a controlled framework for testing how prescribed filament
dynamics map into GONG-like H$\alpha$ observables. Its outputs enable
validation of oscillation-analysis methods, assessment of observational
biases, and generation of labeled datasets for segmentation and
motion-estimation applications.}


\maketitle
\nolinenumbers

\section{Introduction}

Filament dynamics are commonly studied using time sequences of H$\alpha$ observations, which reveal the evolution of their morphology and intensity. However, the observed intensity changes do not provide the three-dimensional motion directly: projection, radiative transfer, background evolution, and instrumental effects all influence the measured signal. Recovering the
underlying plane-of-sky velocity field is therefore an inverse problem.

This inverse problem motivates the development of FILOS as a first step toward the supervised inference of filament motions from H$\alpha$ image sequences. Following the simulation-to-velocity strategy of DeepVel \citep{Asensio...2017} and more recent shallow U-net approaches such as SUVEL \citep{Liu...2025}, a neural network could be trained from paired examples consisting of a synthetic H$\alpha$ image sequence and the corresponding model-defined plane-of-sky velocity field. The velocity field of a real filament is generally unknown, so FILOS addresses this limitation by generating synthetic GONG-like image sequences together with the opacity-weighted velocity fields defined by the model. DeepVel and SUVEL target photospheric flows, whereas FILOS is designed to provide analogous training data for filament dynamics observed in H$\alpha$. The same simulations also provide temporally resolved filament masks, enabling the construction of paired image--mask datasets for training and evaluating segmentation networks. The time--distance and spectral analyses presented below are therefore examples of downstream applications of the synthetic data, rather than the primary purpose of the simulator. 

Generating sufficiently large training sets directly from fully self-consistent simulations of prominence formation and evolution is challenging. Such simulations must combine realistic magnetic-field configurations, thermal and mass-loading processes, non-LTE radiative transfer, and instrumental effects. Moreover, training a neural network requires thousands of diverse examples, making a fully realistic simulation for every case computationally prohibitive. We therefore adopt controlled, simplified parametric models that retain the essential geometric, radiative, instrumental, and kinematic effects while allowing the relevant parameters to be varied systematically. FILOS is intended as a starting point for developing and testing the inference pipeline; its components can later be replaced or refined using more realistic simulations.

Solar prominences are cool, dense plasma structures embedded in the hot solar corona and supported by magnetic fields. When observed against the solar disk, they appear as dark elongated structures and are commonly referred to as filaments. Their morphology, thermodynamic structure, and dynamics are closely linked to the magnetic configuration of the surrounding corona, making them valuable diagnostics of the solar atmosphere \citep{Mackay...2010,GibsonReview...2018}. The interpretation of prominence observations also requires consideration of their complex thermodynamic and radiative properties, particularly because the formation of spectral lines such as H$\alpha$ generally requires non-LTE radiative-transfer modeling \citep{LabrosseReview...2010,HeinzelReview...2025}.

Among the different dynamical phenomena observed in prominences, oscillations are of particular interest because they provide an indirect means of determining physical quantities that are difficult to measure directly such as the geometry and magnetic field of the filament threads. Prominence oscillations have been observed over a broad range of periods, amplitudes, and spatial scales and form the basis of prominence seismology \citep{Oliver...2002,Arregui...2018}. Their properties depend on the restoring forces acting on the prominence plasma and on the geometry of the supporting magnetic field.

This connection is particularly clear for large-amplitude longitudinal oscillations, for which gravity projected along dipped magnetic field lines can provide the dominant restoring force under typical prominence conditions \citep{Luna...2012,LunaModel...2012,Zhang...2012}. In the pendulum approximation, the period is primarily controlled by the radius of curvature of the magnetic dip, with extensions accounting for finite flux-tube cross-sections, nonuniform solar gravity, and noncircular dip geometries \citep{Luna...2016,Luna...2022, Liakh...2020, Liakh...2021}. Measurements of filament oscillations can therefore provide constraints on the geometry and strength of the magnetic field supporting the prominence plasma.

Observational studies have traditionally characterized these motions through visual inspection, manually positioned slits, and time--distance diagrams. These techniques provide measurements of quantities such as the period, displacement amplitude, damping time, and direction of motion, and have been successfully applied to many individual events \citep[see the review by][]{Arregui...2018}. The availability of long-duration H$\alpha$ observations from the Global Oscillation Network Group (GONG) has also enabled larger surveys. Using observations obtained near the maximum of solar cycle 24, \citet{LunaCatalog...2018} compiled a catalog of 196 filament oscillations and found a mean period of approximately 58 min. Their results also showed that both small- and large-amplitude oscillations are common and that a large fraction of the observed motions are predominantly longitudinal.

The size of the available H$\alpha$ archives has motivated the development of automatic alternatives to manual searches. \citet{LunaSpectral...2022} introduced a spectral technique that searches for periodic intensity fluctuations directly in GONG image sequences. \citet{Castello...2025} subsequently accelerated the estimation of spectral backgrounds and significance levels using convolutional neural networks, making the analysis of large image sequences computationally practical. More recently, \citet{Castello...2026} combined automatic filament segmentation with a multi-scale spectral analysis to identify spatially coherent oscillations and determine their location and dominant period.

These developments provide a framework for systematically detecting filament oscillations over large observational archives. A subsequent step is the automatic recovery of additional properties of the detected events, including their displacement amplitude, damping time, onset, polarization with respect to the filament spine, spatial extent, and velocity field. These quantities are required for a more complete description of the dynamics and for their subsequent seismological interpretation. Assessing the accuracy of such measurements from observations is challenging because the three-dimensional filament geometry and the true parameters of the underlying motion are generally unknown. Controlled datasets of synthetic observations in which these quantities are known by construction can provide a direct reference for testing parameter-recovery methods.

Forward modeling provides a natural framework for constructing such synthetic observations. Significant progress has been made in connecting three-dimensional prominence models with observable H$\alpha$ structures. \citet{Gunar...2013} combined magnetic dips obtained from nonlinear force-free-field models with prominence plasma distributed in hydrostatic equilibrium, allowing synthetic spectra to be calculated from physically defined fine structures. This approach was subsequently extended to three-dimensional whole-prominence models containing large ensembles of plasma-filled magnetic structures \citep{Gunar...2015,GunarII...2015}. To efficiently connect these complex models with H$\alpha$ observations, \citet{Heinzel...2015} developed an approximate radiative-transfer method for generating synthetic prominence and filament images.

These forward-modeling studies also demonstrate the importance of observational effects when interpreting prominence structure. In particular, \citet{Gunar...2018} showed that H$\alpha$ visibility and the projection of the intrinsically three-dimensional magnetic and plasma structure can substantially modify the apparent morphology and motion of prominence fine structures. More detailed spectral synthesis has also become possible: \citet{Jenkins...2023} applied 1.5D non-LTE radiative transfer to a three-dimensional MHD filament-prominence simulation, producing synthetic spectra and intensities directly from the simulated plasma. More recently, \citet{Jenkins...2024} extended this approach toward multidimensional radiative transfer, using a 2.5D treatment to investigate the radiative coupling between a prominence and the underlying chromosphere. Synthetic non-LTE spectra have also been calculated from dynamic multidimensional MHD prominence simulations, including H$\alpha$ and several other chromospheric lines \citep{Pietrow...2024}. Together, these developments show the value of translating three-dimensional prominence models into observables when comparing physical models with imaging and spectroscopic measurements.

In this work, we present FILOS as a first step toward generating labeled H$\alpha$ image sequences for learning-based inference of filament motions. FILOS combines the measured morphology of real filaments with an ensemble of three-dimensional dipped threads. A static prominence plasma is distributed along these threads following hydrostatic equilibrium, and its H$\alpha$ opacity is estimated using an extended version of the approximate non-LTE treatment of \citet{Heinzel...2015}. The resulting filament is projected according to its position on the solar disk, degraded to the spatial resolution of GONG, and rendered over observed GONG H$\alpha$ background sequences. The background evolution and observational variability therefore originate from real observations, while the inserted filament geometry is completely specified by the forward model.

Controlled longitudinal and transverse motions can then be imposed on the filament threads, with prescribed amplitude, period, damping time, onset, excitation location, and spatial extent. Longitudinal periods can also be assigned independently to the threads from their curvature and height using the extended pendulum model of \citet{Luna...2022}. Because both the three-dimensional structure and the prescribed dynamics are known throughout the sequence, FILOS provides model-defined reference labels against which quantities recovered from the rendered H$\alpha$ observations can be evaluated. The model also provides derived products such as filament segmentation masks and projected velocity fields, extending its use to the development and validation of image-processing and machine-learning methods.

The main objective of FILOS is therefore to provide paired synthetic H$\alpha$ image sequences, model-defined velocity fields, and segmentation masks for developing and evaluating motion-inference and image-segmentation methods. In this paper, we describe the forward model and illustrate its capabilities using a representative longitudinal oscillation. We show how the imposed dynamics appear after projection and observational degradation, compare parameters recovered from a conventional time--distance analysis with their prescribed values, and present additional model-defined reference products generated directly by the model.

The paper is organized as follows. In Sect.~2, we describe the FILOS forward model, including the construction of the three-dimensional filament geometry, the plasma and H$\alpha$ opacity model, the generation of synthetic observations, and the imposed oscillatory dynamics. In Sect.~3, we present a representative FILOS simulation, analyze the resulting synthetic oscillation, and illustrate additional ground-truth products generated by the model. Finally, Sect.~4 summarizes the main conclusions.

\section{FILOS forward model}
\label{sec:forward_model}

FILOS is a parametric forward model for generating synthetic H$\alpha$ observations of oscillating solar filaments. The model combines the measured morphology of a real filament with a prescribed three-dimensional thread structure, a static thermodynamic plasma with an approximate H$\alpha$ opacity, and controlled kinematic motions. The resulting filament is projected onto the image plane and rendered over real GONG H$\alpha$ observations.

The purpose of FILOS is to construct controlled observations rather than to simulate the physical evolution of a filament. The three-dimensional geometry is prescribed directly, the plasma properties are calculated from static equilibrium assumptions and tabulated quantities, and the oscillatory motions are imposed analytically. The geometry and dynamics of every synthetic sequence are therefore known by construction.

\subsection{Three-dimensional filament geometry}
\label{sec:geometry}

The large-scale morphology of each synthetic filament is defined by a measured
planar spine $\mathbf{C}(u)$, where $u$ denotes arclength. At each position
along the spine, its local tangent $\mathbf{t}(u)$ and horizontal normal
$\mathbf{n}(u)=\hat{\mathbf{z}}\times\mathbf{t}(u)$ define a reference frame
for positioning and orienting the individual threads.

\begin{figure}
    \centering
    \includegraphics[width=\columnwidth]{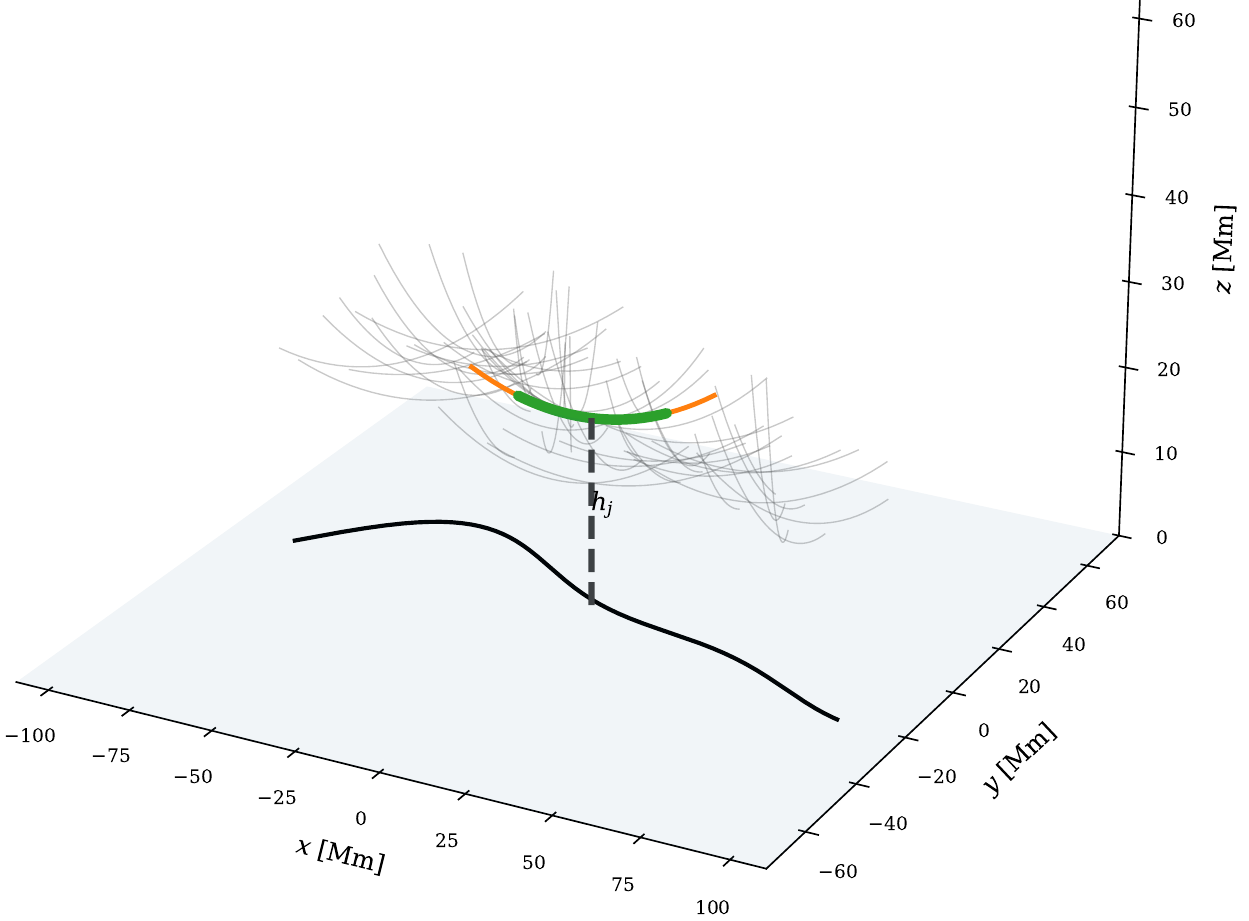}
    \caption{
    Schematic three-dimensional structure of a FILOS filament. The measured
    filament spine is shown on the solar-surface plane ($z=0$) and is populated
    by an ensemble of dipped filament threads. One representative thread is
    highlighted together with its minimum height $h_j$ and central
    plasma-loaded region. For clarity, only a subset of the threads used in a
    typical realization is shown.
    }
    \label{fig:filament_3d}
\end{figure}

The observed shape and physical scale of the spine are preserved, although its position and orientation within the field of view can vary between realizations. An ensemble of threads is distributed within a smoothly varying channel around the spine. Each thread is modeled with a circular cross section whose radius is drawn from a configurable range and decreases smoothly toward both endpoints to a prescribed minimum fraction of its central value. Figure~\ref{fig:filament_3d} shows the resulting three-dimensional structure.

A prescribed linear thread density \(\rho_{\rm th}\) sets the target population as \(N_{\rm th}\simeq\rho_{\rm th}L_{\rm spine}\), subject to a configurable upper limit. The ensemble is then constructed sequentially. For each thread, an anchor position is selected along the spine and displaced laterally within the local channel envelope. The shear angle is drawn from a prescribed distribution about a mean value, with its sign determined by the selected filament chirality. The dip-bottom height, geometrical length, curvature radius, and cross-section radius are likewise sampled from configurable distributions, subject to the geometrical constraints described below. During placement, a minimum three-dimensional centerline-separation criterion limits excessive clustering. The procedure is repeated until the target thread population is reached.

\begin{figure*}
    \sidecaption
    \includegraphics[width=5.9cm]{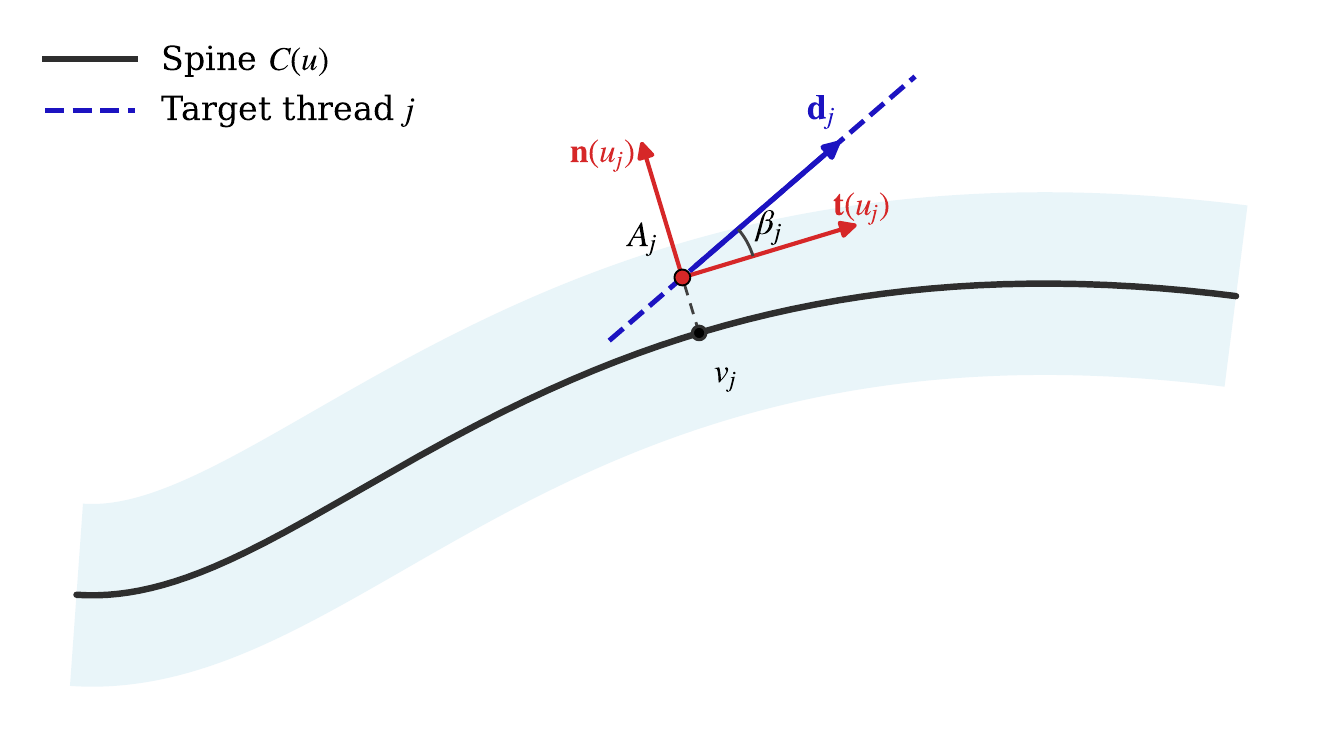}%
    \hspace{0.2cm}%
    \includegraphics[width=5.9cm]{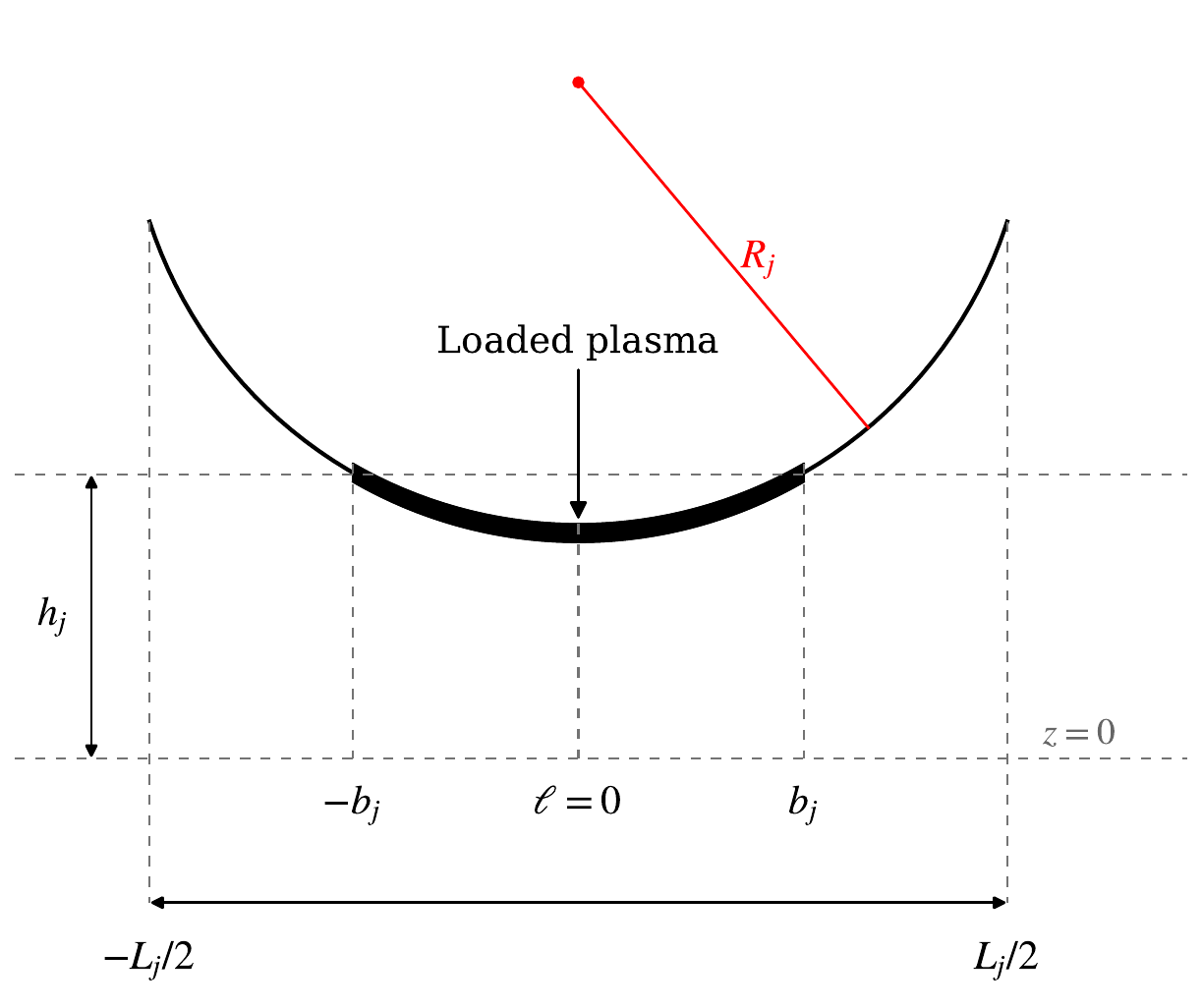}
    \caption{Geometrical construction of the filament threads.
    \textit{Left:} Horizontal distribution of threads around the
    measured spine $\mathbf{C}(u)$. The local tangent
    $\mathbf{t}(u_j)$ and normal $\mathbf{n}(u_j)$ define the
    reference frame at position $u_j$. The anchor
    $\mathbf{A}_j$ is displaced by $v_j$ from the spine,
    while the shear angle $\beta_j$ determines the thread
    direction $\mathbf{d}_j$.
    \textit{Right:} Vertical geometry of a thread following
    a circular arc of curvature radius $R_j$ and minimum
    height $h_j$. The complete thread extends over
    $[-L_j/2,L_j/2]$, while cool plasma occupies
    $[-b_j,b_j]$.
    }
    \label{fig:thread_geometry}
\end{figure*}

For thread $j$, the anchor position and horizontal direction are
\begin{align}
\mathbf{A}_j &=
\mathbf{C}(u_j)+v_j\mathbf{n}(u_j), \\
\mathbf{d}_j &=
\cos\beta_j\,\mathbf{t}(u_j)
+
\sin\beta_j\,\mathbf{n}(u_j),
\label{eq:thread_placement}
\end{align}
where $v_j$ is the lateral displacement from the spine and $\beta_j$ is the
shear angle. The left panel of Fig.~\ref{fig:thread_geometry} illustrates this
construction.

The thread is then extended into three dimensions by assigning a circular-segment dip
in the vertical plane containing $\mathbf{d}_j$. Its centerline is
\begin{equation}
\mathbf{r}_j(\ell)
=
\mathbf{A}_j
+
R_j
\sin\left(\frac{\ell}{R_j}\right)
\mathbf{d}_j
+
\left[
h_j
+
R_j
\left(
1-\cos\frac{\ell}{R_j}
\right)
\right]
\hat{\mathbf{z}},
\qquad
|\ell|
\leq
\frac{L_j}{2},
\label{eq:thread_geometry}
\end{equation}
where $\ell$ is the field-line arc-length coordinate measured from the bottom of the dip, $R_j$ is the curvature radius, $L_j$ is the geometrical thread length, and $h_j$ is its minimum height above the solar surface. We require $L_j\leq\pi R_j$, so that the prescribed arc does not exceed a semicircle. Cool filament plasma occupies only the central interval $|\ell|\leq b_j$, as illustrated in the right panel of Fig.~\ref{fig:thread_geometry}.

The circular geometry provides a simple constant-curvature representation consistent with the pendulum model for longitudinal prominence oscillations \citep{LunaModel...2012,Luna...2016}. More general dip geometries, such as semi-elliptical or sinusoidal profiles, have spatially varying curvature, but their longitudinal oscillation periods are still determined primarily by the local radius of curvature near the dip bottom \citep{Luna...2022}. For sufficiently large displacements, however, the plasma can sample regions farther from the dip bottom where the curvature differs, so more general dip geometries may affect the detailed nonlinear dynamics.

The measured spine therefore controls the large-scale filament morphology, while the sampled and geometrically constrained thread properties determine its internal three-dimensional structure.

\subsection{Static plasma loading and H\texorpdfstring{$\alpha$}{alpha} opacity}
\label{sec:plasma_opacity}

Once the geometry has been constructed, a static thermodynamic structure is assigned to each thread within the symmetric interval $|\ell|\leq b_j=q_jL_j/2$, as shown in Fig.~\ref{fig:thread_geometry}b. Here, $q_j$ is the fraction of the geometrical thread length occupied by the loaded plasma, with $0<q_j\leq1$. At each position \(\ell\), the thermodynamic quantities are assumed to be uniform across the local circular cross section, so the plasma structure is described only along the field-line coordinate \(\ell\). This interval contains a cool central core and its
prominence--corona transition region.

Following the prominence-thread formulation of \citet{Gunar...2013}, the temperature increases from $T_{\rm c}$ at the dip bottom to $T_{\rm tr}$ at the loaded boundaries:
\begin{equation}
T_j(\ell) = T_{\rm c} + (T_{\rm tr}-T_{\rm c}) \left(\frac{|\ell|}{b_j}\right)^\gamma,
\qquad |\ell|\leq b_j,
\label{eq:temperature_profile}
\end{equation}
where $\gamma$ controls the steepness of the temperature profile.
For a given loaded fraction, the pressure is obtained by integrating one-dimensional hydrostatic balance along the circular thread,
\begin{equation}
\frac{{\rm d}p_j}{{\rm d}\ell} = -\rho_j g_0 \frac{{\rm d}z_j}{{\rm d}\ell},
\label{eq:hydrostatic_balance}
\end{equation}
from the boundaries toward the dip bottom, with
$g_0=274~{\rm m\,s^{-2}}$ and $p_j(\pm b_j)=p_{\rm tr}$.

For this hydrostatic calculation, the hydrogen ionization fraction is approximated as
\begin{equation}
i_{\rm hyd}(T) = 1-(1-i_{\rm c}) \left(\frac{T_{\rm tr}-T}{T_{\rm tr}-T_{\rm c}}\right)^2,
\end{equation}
with $i_{\rm c}=0.3$. Assuming neutral helium and a helium-to-hydrogen number abundance $\alpha = 0.1$, the dimensionless mean molecular weight and the mass density are
\begin{equation}
\begin{aligned}
\mu_{\rm m}(T) &= \frac{1 + 4\alpha}{1 + \alpha + i_{\rm hyd}(T)} = \frac{1.4}{1.1 + i_{\rm hyd}(T)},\\
\rho_j &= \frac{\mu_{\rm m}(T_j)m_{\rm H}p_j}{k_{\rm B}T_j},
\end{aligned}
\end{equation}
where $m_{\rm H}$ is the hydrogen mass and $k_{\rm B}$ is the
Boltzmann constant.

The loaded fraction $q_j$ is determined by a prescribed target field-aligned column mass $M_\star$, common to all threads. For thread $j$, the column mass is
\begin{equation}
M_j(q_j)=2\int_0^{q_jL_j/2}\rho_j(\ell;q_j)\,{\rm d}\ell.
\label{eq:column_mass}
\end{equation}
The temperature, pressure, and density profiles are recomputed as $q_j$ is varied. We adopt $M_\star=10^{-4}~{\rm g\,cm^{-2}}$ and $p_{\rm tr}=0.015~{\rm dyn\,cm^{-2}}$. If the full-dip column mass $M_j(1)$ exceeds $M_\star$, we determine $q_j$ numerically from $M_j(q_j)=M_\star$. Otherwise, we set $q_j=1$, and the realized column mass is $M_j(1)$. Since the integration follows the thread centerline, $M_j$ represents a field-aligned
rather than a line-of-sight column mass.

The local H$\alpha$ opacity is estimated using a calibrated extension of the approximate non-LTE treatment of \citet{Heinzel...2015}. The response table provides the hydrogen ionization degree $i=n_{\rm p}/n_{\rm H}$ and the population factor $F=n_{\rm p}^2/n_2$ as functions of local height, temperature, and pressure. Here, $n_{\rm H}$ is the total hydrogen number density, $n_{\rm p}$ is the proton density, and $n_2$ is the population density of the second hydrogen level.

The extended height dependence was obtained from non-LTE calculations with Promweaver \citep{Osborne...2022}, based on the Lightweaver framework \citep{Osborne...2021}, and calibrated multiplicatively to reproduce the reference values reported by \citet{Heinzel...2015} at 10, 20, and 30~Mm. The resulting table spans heights of 1--100~Mm, temperatures of 6000, 8000, 10000, 12000, and 14000~K, and pressures of 0.01, 0.02, 0.05, 0.10, and 0.20~dyn\,cm$^{-2}$. The quantities $i$ and $F$ are evaluated by linear interpolation on this three-dimensional grid. Height and pressure queries outside the tabulated ranges are clipped to the nearest boundary for the lookup. Loaded plasma outside the supported temperature interval remains part of the hydrostatic structure but is assigned zero H$\alpha$ opacity.

Under the neutral-helium assumption, $n_{\rm e}=n_{\rm p}$, so the local pressure and temperature give
\begin{equation}
n_{\rm e}=\frac{i\,p_j}{k_{\rm B}T_j(1.1+i)},\qquad n_2=\frac{n_{\rm e}^2}{F}.
\end{equation}
The tabulated ionization degree is used only for the opacity
calculation; the hydrostatic structure remains that obtained with
$i_{\rm hyd}$.

The line-center absorption coefficient is
\begin{equation}
\kappa_{0,j}=1.7\times10^{-2}\frac{n_2}{\sqrt{\pi}\Delta\nu_{\rm D}},
\label{eq:halpha_opacity}
\end{equation}
with $n_2$ in ${\rm cm^{-3}}$, $\Delta\nu_{\rm D}$ in Hz, and
$\kappa_{0,j}$ in ${\rm cm^{-1}}$. The Doppler width includes thermal
broadening and a microturbulent velocity of
$v_{\rm turb}=5~{\rm km\,s^{-1}}$. We use
$\lambda_0=6562.8$~\AA.

For a thread element with local radius $r_j(\ell)$, the line-center
optical-depth scale across its diameter is
\begin{equation}
\tau_{0,j}(\ell)=2\,r_j(\ell)\,\kappa_{0,j}(\ell),
\end{equation}
with $r_j$ expressed in cm. These local contributions are subsequently
rasterized on the oversampled image grid as described in
Sect.~\ref{sec:image_formation}.

\subsection{Projection and synthetic \texorpdfstring{H$\alpha$}{Halpha} images}
\label{sec:image_formation}

The three-dimensional filament and its H$\alpha$ opacity are converted into a GONG-like observation by projection onto the image plane, construction of an optical-depth map, spatial degradation, and combination with a real H$\alpha$ background.

The projection is determined by the location of the selected observation on the solar disk. Let $\mu=\cos\vartheta$ be the viewing cosine, $\mathbf{e}$ the horizontal unit vector toward the nearest limb, and $\mathbf{c}$ the center of the field of view. A point with horizontal position $\mathbf{r}_{\rm h}$ and height $z$ is projected as
\begin{equation}
\Pi(\mathbf{r}_{\rm h},z)=
\mathbf{r}_{\rm h}+(\mu-1)
\left[(\mathbf{r}_{\rm h}-\mathbf{c})\cdot\mathbf{e}\right]\mathbf{e}
+
\sqrt{1-\mu^2}\,z\mathbf{e}.
\label{eq:projection}
\end{equation}
The second term describes horizontal foreshortening, while the third accounts for the apparent displacement produced by the filament height. The viewing geometry is obtained from the selected GONG observation and is assumed constant across the cropped field of view.

The absorbing thread elements are rasterized onto an internally oversampled image grid. Each projected centerline element is represented by an isotropic Gaussian kernel with standard deviation $\sigma_j(\ell)=r_j(\ell)/2$. The kernels are integrated over the image pixels, with weights determined by the local transverse optical depth and the arclength contribution of each element. An additional smooth envelope reduces these contributions near the two ends of the filament spine. Contributions from all elements are summed in optical depth, including projected thread overlaps, to form the map $\tau(\mathbf{X})$.

The corresponding line-center transmission is
\begin{equation}
T_{\rm line}(\mathbf{X})=\exp[-\tau(\mathbf{X})].
\label{eq:line_transmission}
\end{equation}
To approximate the spatial response of GONG, this transmission map is convolved with a Gaussian point-spread function and averaged in blocks to the native pixel scale, yielding $T_{\rm nat}$. The degradation is applied to the synthetic transmission, while the observed background is retained at its native resolution because it already contains the instrumental response and observational noise.

The finite spectral width of the GONG H$\alpha$ filter is represented approximately through a fixed line-contribution fraction $\eta$. The effective transmission and synthetic intensity are
\begin{align}
T_{\rm obs}&=(1-\eta)+\eta T_{\rm nat}, \\
I_{\rm syn}&=B\,T_{\rm obs}+S(1-T_{\rm obs}),
\label{eq:synthetic_intensity}
\end{align}
where $T_{\rm nat}$ is the spatially degraded line-center transmission, $B$ is the observed GONG background, and $S$ is a spatially uniform effective filament source function. The two intensity terms represent the transmitted background and the filament source contribution, respectively. We use $\eta=0.65$ in the reference configuration. This fixed-fraction approximation does not account for velocity-dependent changes in visibility caused by Doppler shifts through the GONG H$\alpha$ passband.

The validity of this approximation depends on the line-of-sight
velocity. In the reference experiment
(Sect.~\ref{subsec:longitudinal}), the nominal amplitude
of 15~Mm, median period of 60.8~min, and viewing angle
($\mu=0.883$) give an upper line-of-sight velocity estimate
of 12.1~km\,s$^{-1}$ ($\Delta\lambda\simeq0.27$~\AA)
for a thread with the median period and maximum line-of-sight
projection, before spatial amplitude weighting.

Given the GONG H$\alpha$ filter bandwidth of
0.5--0.6~\AA\footnote{\url{https://nso.edu/data/nisp-data/h-alpha/}},
Doppler-dependent visibility changes may be significant.
The fixed-fraction approximation is therefore most appropriate
for small line-of-sight velocities, particularly near disk center
or for motions predominantly perpendicular to the line of sight.
A quantitative treatment of Doppler effects would require
the spectral line profile and instrumental transmission function.

The source-to-background intensity ratio is determined from the values reported by \citet{Heinzel...2015} at 10, 20, and 30~Mm, using the mean dip-bottom height of the filament. Within this height range, these values are linearly interpolated. At heights below 10~Mm or above 30~Mm, the corresponding endpoint value is extended using the geometrical dilution factor of the incident solar radiation,
\begin{equation}
W(H)=\frac{1}{2}
\left[1-\sqrt{1-\frac{R_\odot^2}{(R_\odot+H)^2}}
\right],
\end{equation}
such that the source-function ratio scales as $W(H)/W(H_{\rm a})$, where $H_{\rm a}$ is the nearest reference anchor height. The resulting ratio is multiplied by the median intensity of the initial background to obtain $S$, which is shared by all threads and held fixed throughout the sequence.

The resulting images combine the prescribed filament structure and opacity with the chromospheric features, observational noise, and temporal variability present in the real GONG background.

\subsection{Prescribed kinematic oscillations}
\label{sec:oscillations}

A time-dependent FILOS sequence is produced by prescribing the displacement of the previously constructed threads. The thermodynamic and opacity profiles are computed for the initial configuration and remain attached to their corresponding threads throughout the sequence. The temporal evolution is therefore purely kinematic.

For longitudinal oscillations, the period can be specified directly or assigned from the curvature and height of each thread using the pendulum model of \citet{LunaModel...2012, Luna...2022}. Here, we use a generalized form of these models that includes the height dependence of solar gravity and the curvature of the solar surface.
\begin{equation}
P_{\parallel,j}
=
2\pi
\left[
g_0
\left(
\frac{R_\odot}{R_\odot+h_j}
\right)^2
\left(
\frac{1}{R_j}
+
\frac{1}{R_\odot+h_j}
\right)
\right]^{-1/2}.
\label{eq:luna_period}
\end{equation}
This expression is used only to assign the oscillation period. Differences in $R_j$ and $h_j$ consequently produce a distribution of longitudinal periods within the same synthetic filament. Transverse periods can be prescribed independently.

The oscillatory region is localized around a selected excitation position. For each thread, we calculate the initial three-dimensional distance $d_j$ between its optical-depth-weighted material centroid and the excitation site. Its oscillation amplitude is then multiplied by the spatial weight
\begin{equation}
w_j
=
\exp
\left[
-\ln 2
\left(
\frac{d_j}{L_\star}
\right)^{1.1}
\right],
\label{eq:spatial_weight}
\end{equation}
where $L_\star$ is the distance at which the amplitude decreases to half of its central value. This weight is calculated once from the initial configuration and remains fixed throughout the sequence. It therefore controls the spatial coherence and extent of the imposed oscillation.

For simplicity, we prescribe two coherent motion components relative
to each thread: a longitudinal component along its horizontal
direction $\mathbf{d}_j$ and a transverse component along
$\hat{\mathbf{z}}\times\mathbf{d}_j$. Both components produce rigid
translations of the complete curved thread, with its plasma and
opacity distribution transported together. The longitudinal
component therefore represents a horizontal displacement aligned
with the thread rather than plasma motion along a fixed magnetic dip.

For either component $c\in\{\parallel,\perp\}$, the imposed
displacement is
\begin{equation}
\xi_{c,j}(t)=A_c\exp\left[-\frac{t-t_0}{\tau_{\rm d}}\right]
\left[
\sin\left(\frac{2\pi(t-t_0)}{P_{c,j}}+\phi\right)
-\sin\phi
\right],
\qquad t\geq t_0,
\label{eq:damped_motion}
\end{equation}
with $\xi_{c,j}(t)=0$ for $t<t_0$.

Here, $A_c$ is the displacement amplitude scale,
$\tau_{\rm d}$ is the damping time, $t_0$ is the excitation
time, and $\phi$ is the initial phase. The coherent displacement
of thread $j$ is then
\begin{equation}
\mathbf{D}^{\rm coh}_j(t)
=
w_j\left[
\xi_{\parallel,j}(t)\mathbf{d}_j+
\xi_{\perp,j}(t)(\hat{\mathbf{z}}\times\mathbf{d}_j)
\right].
\label{eq:coherent_displacement}
\end{equation}

Because each thread is inclined by an angle $\beta_j$ relative
to the local filament spine, its longitudinal displacement has
both spine-parallel and spine-perpendicular components,
proportional to $\cos\beta_j$ and $\sin\beta_j$, respectively.
Consequently, a purely longitudinal thread motion can produce
an apparent transverse displacement relative to the observed
filament spine. The relative contributions are further modified
by projection onto the image plane. The longitudinal and
transverse labels in FILOS therefore refer to the prescribed
thread-relative directions, while the polarization inferred
from observations also depends on the thread orientation and
viewing geometry. This distinction is relevant to the interpretation of observed
filament oscillations, for which stereoscopic and Doppler
measurements can help constrain the actual direction of
motion \citep{ShenWinking...2014,Tan...2023}.

An independent isotropic random walk can optionally be added to each thread to represent small-scale stochastic motions. At each time step, its direction is sampled randomly and its step length is drawn from a prescribed interval. This component is added to the coherent displacement described above.

At every frame, the displaced thread geometry is projected again onto the image plane, a new optical-depth map is constructed, and Eq.~(\ref{eq:synthetic_intensity}) is evaluated using the corresponding GONG background frame. The thermodynamic and opacity profiles are transported with the threads rather than recalculated from a time-dependent plasma evolution.

FILOS therefore produces controlled synthetic H$\alpha$ sequences in which the filament morphology, oscillation period, amplitude, damping, excitation position, and spatial coherence are specified and known. The combination of these controlled filament properties with real GONG backgrounds provides realistic observational sequences for testing and characterizing filament-oscillation analysis methods.

\section{Results}
In this section, we illustrate the output of FILOS using representative synthetic filament oscillations. We first consider a reference simulation to show how the prescribed three-dimensional structure and dynamics appear in the resulting GONG-like H$\alpha$ observations. We then use the known model configuration and derived products to examine how these synthetic sequences can be analyzed with methods commonly applied to observational data.

\subsection{Longitudinal oscillation example} \label{subsec:longitudinal}

We consider a reference simulation of a longitudinal filament oscillation, with the main configuration parameters summarized in \Cref{tab:reference_simulation}. The filament is constructed from 5392 individual threads distributed along a 154.1 Mm spine. The threads have different lengths, shear angles, and curvature radii, producing a three-dimensional structure with a range of local geometries. Longitudinal motions are prescribed using the pendulum model, with the period of each thread determined by its curvature radius and height. For this experiment, the resulting thread periods range from 44.2 to 87.6 min, with a median of 60.8 min. The oscillation begins at $t_0=20$ min, with a prescribed longitudinal amplitude scale of 15 Mm and a damping time of 15.1 h. The oscillating structure corresponds to the rightmost teardrop-shaped substructure of the filament.

\begin{table}[!ht]
\caption{Main parameters of the reference FILOS simulation.}
\label{tab:reference_simulation}
\centering
\small
\begin{tabular}{@{}ll@{}}
\hline\hline
Parameter & Value \\
\hline
\multicolumn{2}{l}{\textit{Filament geometry}} \\
Spine length              & 154.1 Mm \\
Spine width               & 7.5 Mm \\
Number of threads         & 5392 \\
Thread density            & 35 Mm$^{-1}$ \\
Thread length ($L_j$)     & 20--100 Mm \\
Thread shear angle ($\beta_j$) & $25 \pm 6^\circ$ \\
Prominence height         & $20 \pm 2$ Mm \\
Thread curvature radius ($R_j$) & 50--250 Mm \\
                          & (median 100 Mm) \\
Thread cross-section radius & 75--300 km \\
Endpoint-to-central radius ratio & 0.4 \\
\hline
\multicolumn{2}{l}{\textit{Plasma properties}} \\
Column mass               & $10^{-4}$ g\,cm$^{-2}$ \\
Core temperature          & 7000 K \\
PCTR temperature          & $10^{5}$ K \\
Microturbulent velocity   & 5 km\,s$^{-1}$ \\
\hline
\multicolumn{2}{l}{\textit{Oscillation}} \\
Oscillation model         & Pendulum model \\
Longitudinal amplitude scale ($A_\parallel$) & 15 Mm \\
Transverse amplitude ($A_\perp$) & 0 Mm \\
Longitudinal period ($P_\parallel$) & 44.2--87.6 min \\
                          & (median 60.8 min) \\
Damping time ($\tau$)     & 15.1 h \\
Oscillation onset ($t_0$) & 20 min \\
Half-strength distance ($L_\star$) & 10 Mm \\
Initial phase ($\phi$) & 0 rad \\
\hline
\multicolumn{2}{l}{\textit{Synthetic observation}} \\
Cadence                   & 60 s \\
Number of frames          & 220 \\
Sequence duration         & 219 min \\
Disk position             & $\mu = 0.883$ \\
\hline
\end{tabular}
\end{table}

\begin{figure*}[!h]
    \sidecaption
    \includegraphics[width=12cm]{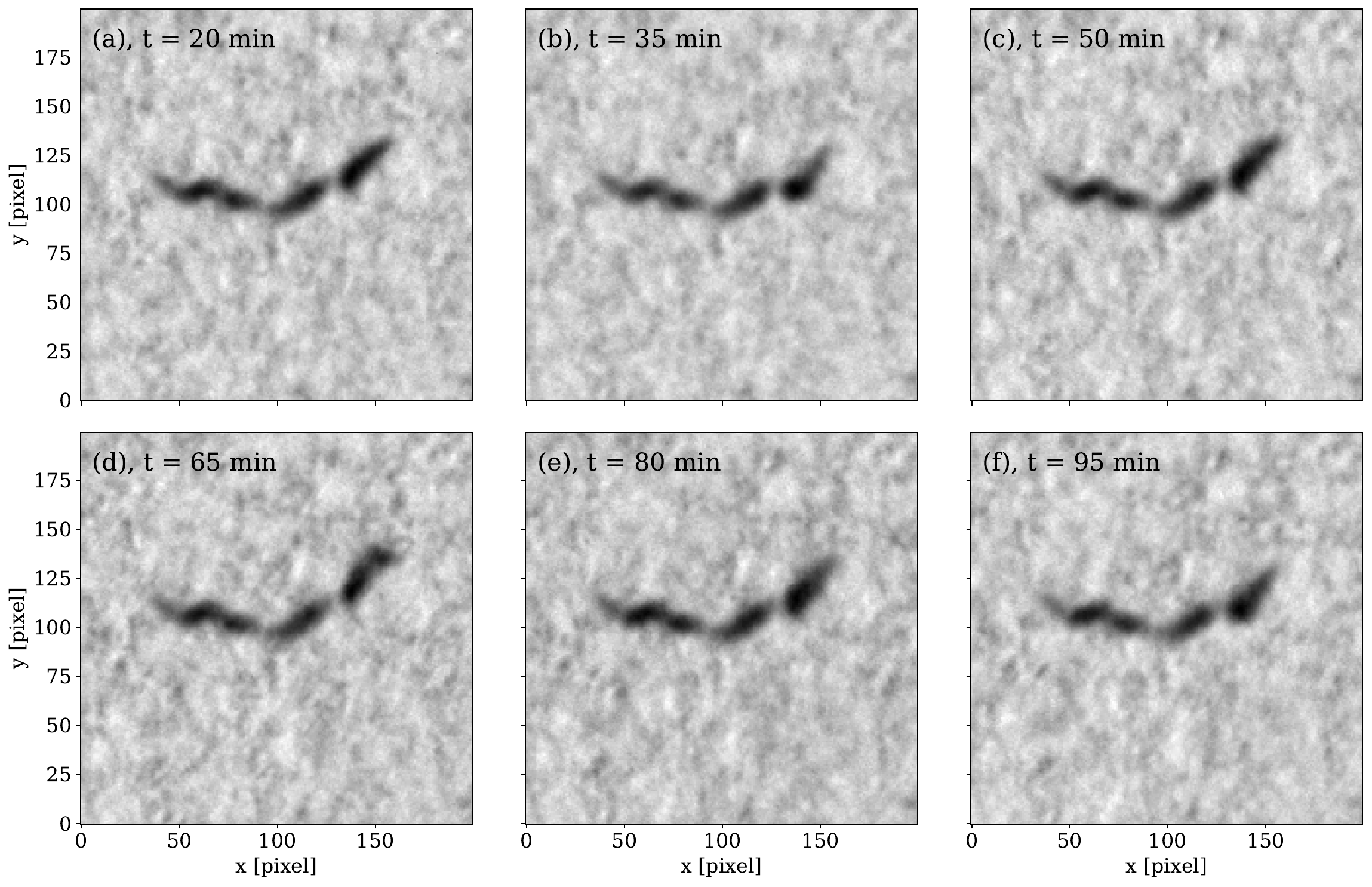}
    \caption{Representative frames of a FILOS-generated
    longitudinal filament oscillation. The sequence has a
    cadence of 1 min, and the displayed frames are separated
    by 15 min. The associated movie is available online.}
    \label{fig:video_frames}
\end{figure*}
Figure~\ref{fig:video_frames} shows representative H$\alpha$ frames from the synthetic sequence. Although the sequence has a cadence of 1 min, the frames shown here are separated by 15 min to sample different phases of the motion. The prescribed longitudinal oscillation appears as a displacement of the absorbing filament material along the structure, while the surrounding chromospheric pattern remains that of the original GONG background. Panel (a) corresponds to the oscillation onset and shows the filament at its initial equilibrium position. Panels (b)--(f) show the subsequent evolution at 15 min intervals.

\begin{figure*}[!ht]
    \centering
    \includegraphics[width=\linewidth]{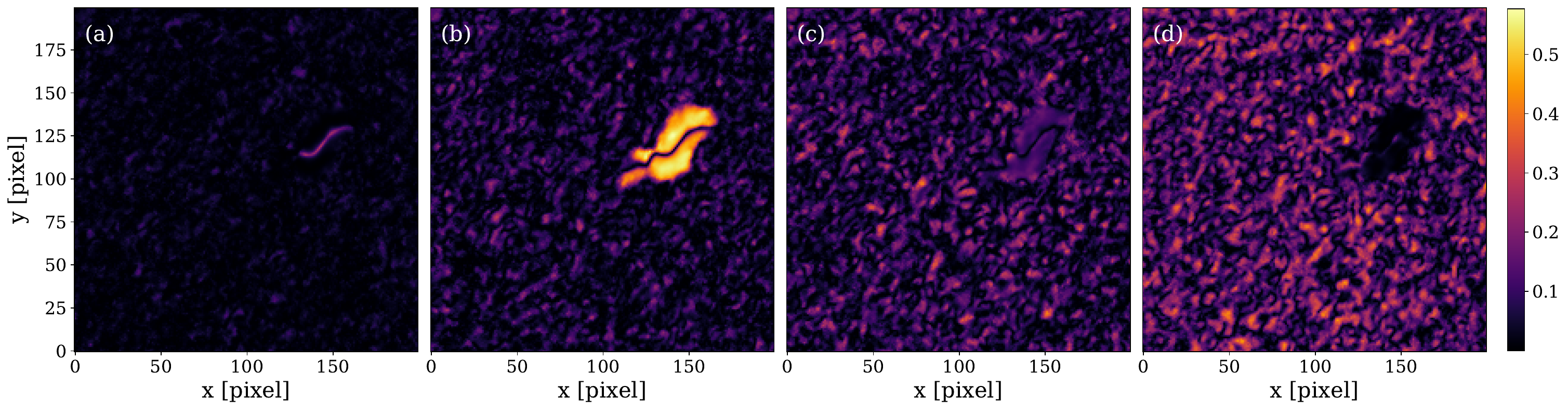}
    \caption{Spatial distribution of the mean standard-normalized Lomb–Scargle power in the period intervals (a) [25--35] min, (b) [55--65] min, (c) [85--95] min, and (d) [115--125] min for the reference longitudinal oscillation. The color scale is the dimensionless standard-normalized Lomb–Scargle power.}
    \label{fig:psd_maps}
\end{figure*}

The synthetic sequence can also be examined in the frequency domain using the same type of spectral analysis applied to observational H$\alpha$ sequences \citep[see][]{LunaSpectral...2022, Castello...2025, Castello...2026}. Figure~\ref{fig:psd_maps} shows the spatial distribution of the standard-normalized Lomb–Scargle power averaged over selected period ranges. As expected, \cref{fig:psd_maps}(b), which contains the period range around the median thread period, shows enhanced power throughout the region swept by the oscillating plasma. The narrow feature in \cref{fig:psd_maps}(a) is consistent with the frequency-doubling, or period-halving, effect. Pixels near the equilibrium position are crossed by the oscillating absorbing structure twice during each displacement cycle, producing an intensity modulation at approximately half the oscillation period. This feature occupies the central region between the two main high-power areas in panel (b).

The period-halving signature in \Cref{fig:psd_maps}(a) arises from the geometrical passage of the absorbing structure through its equilibrium position twice per oscillation cycle. In real H$\alpha$ observations, the measured intensity can also depend on Doppler shifts of the absorption profile relative to the instrumental passband, as illustrated by observations of winking filaments \citep{ShenWinking...2014, ShenChain...2014}. Such effects may modify the spatial distribution and relative strength of the observed periodic signals.

\begin{figure*}
    \sidecaption
    \includegraphics[width=4.5cm]{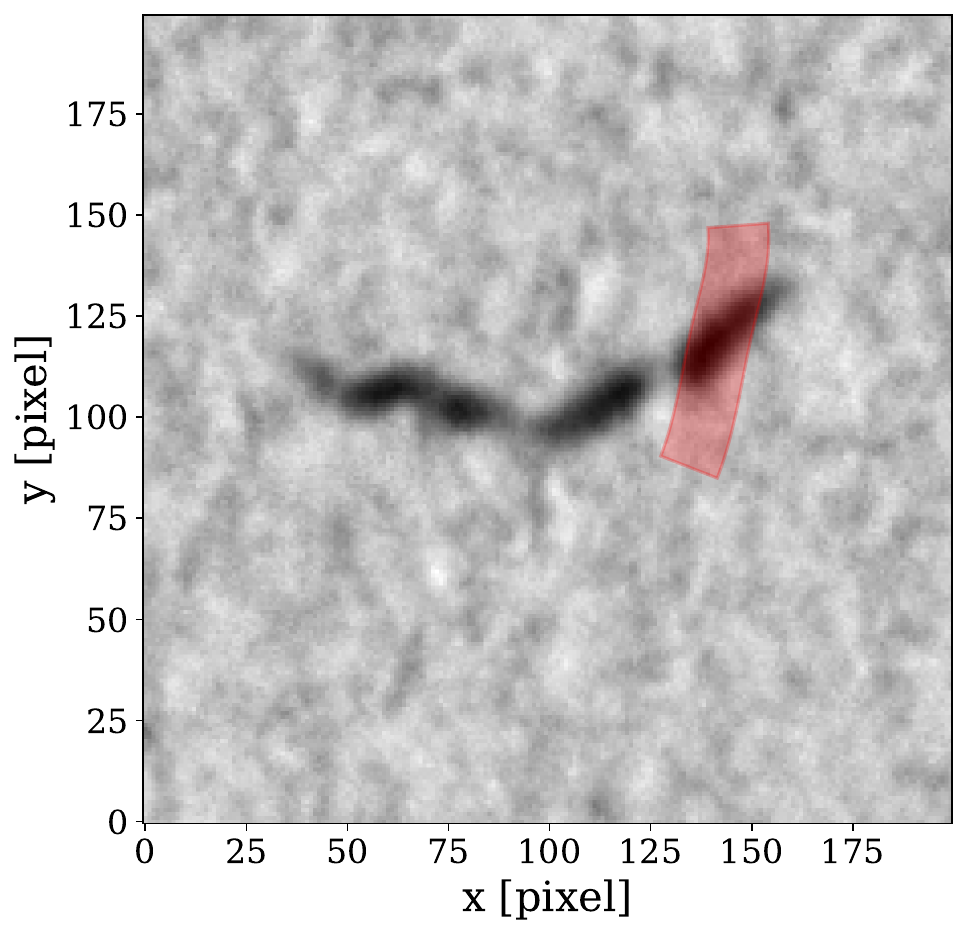}
    \hspace{0.1cm}
    \includegraphics[width=7.4cm]{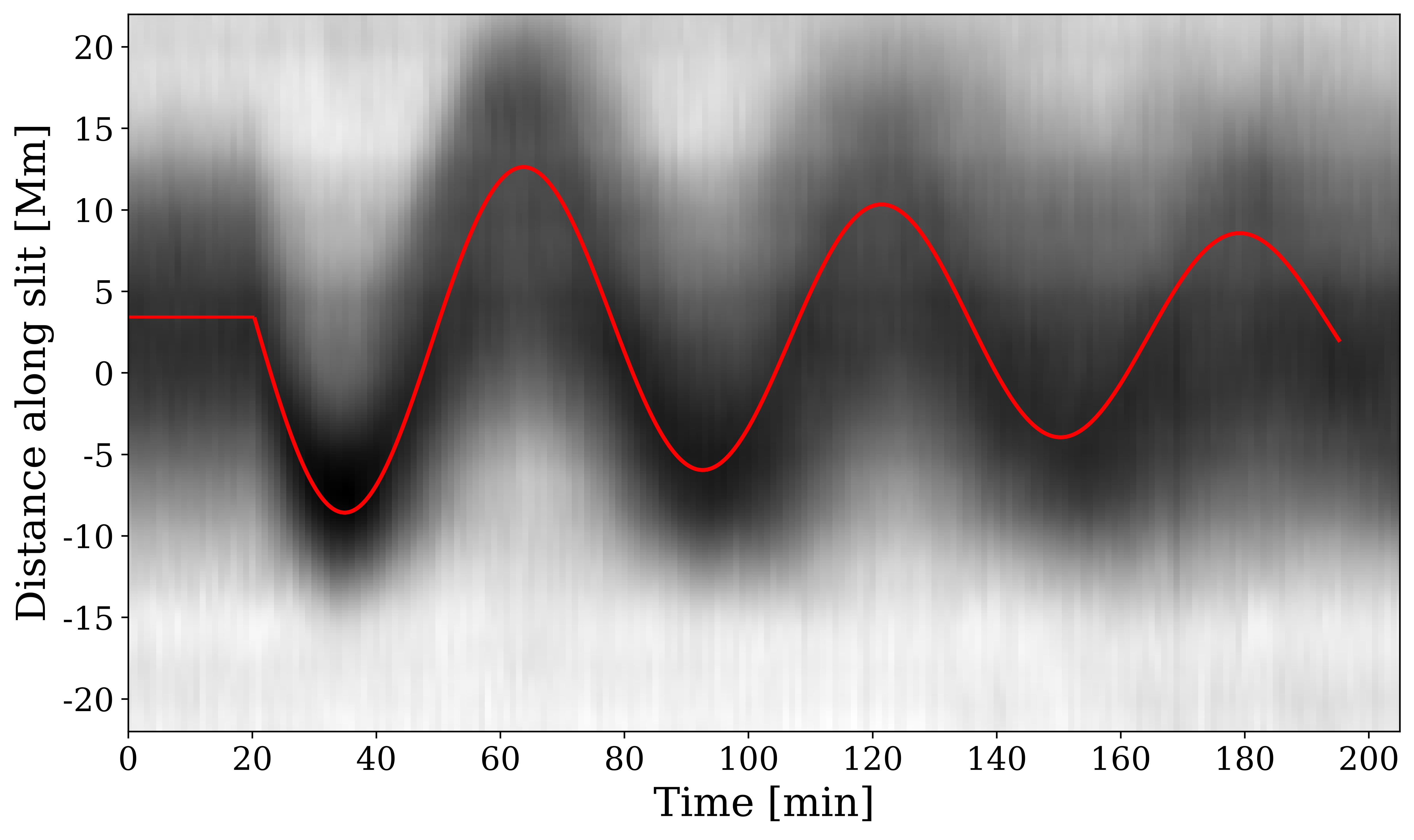}
    \caption{Time--distance analysis of the reference experiment. Left:
    first H$\alpha$ frame of the experiment described in
    \cref{subsec:longitudinal}, with the slit used to track the filament
    motion indicated by the red box. Right: corresponding time--distance
    diagram, with the best-fit damped oscillation model shown in red.}
    \label{fig:slit_fit}
\end{figure*}

To further characterize the synthetic oscillation, we applied a conventional time--distance analysis. A slit was placed across the region swept by the filament during its motion, as shown in \cref{fig:slit_fit}. The resulting time--distance diagram traces the collective displacement of the absorbing structure.
The time--distance diagram was constructed using a slit of 11 pixels across the oscillating structure, averaging the intensity across the slit width at each position. The dark absorption band was tracked manually. The resulting trajectory was fitted with a decaying sinusoidal model
\begin{equation}
x(t) =
\begin{cases}
x_{\rm eq}, & t < t_0, \\[6pt]
x_{\rm eq} + 
A \exp \left[ -\frac{t-t_0}{\tau} \right]
\left[\sin\left(\frac{2\pi(t-t_0)}P+\phi\right)\right],
& t \geq t_0.
\end{cases}
\label{eq:damped_oscillation_fit}
\end{equation}
Here, \(x_{\rm eq}\) denotes the equilibrium position, $A$ is the amplitude, $t_0$ is the onset time, $\tau$ is the damping time, $P$ is the period and $\phi$ is the initial phase.

The fit includes the tracked positions before and after the oscillation onset. The initial phase is fixed at \(\phi=0\), consistent with the reference simulation, while \(x_{\rm eq}\), \(A\), \(t_0\), \(P\), and \(\tau\) are free parameters. The pre-onset data constrain the stationary equilibrium trajectory, while the transition to oscillatory motion determines \(t_0\).

The fitted parameters obtained are summarized together with the relevant FILOS inputs in \Cref{tab:oscillation_fit}. The quoted parameter uncertainties correspond to one-standard-deviation uncertainties derived from the covariance matrix of the fit. The equilibrium position $x_{eq}$ depends on the arbitrary origin of the slit coordinate and is therefore omitted from the comparison.

The recovered onset time, $19.53\pm0.23$ min, closely matches the prescribed value of 20 min. The fitted period is $57.71\pm0.35$ min, which lies within the expected thread-period range of 44.2--87.6 min and is close to its 60.8 min median. This agreement should be interpreted at the level of the collective motion: the fitted value characterizes the dominant periodicity of the H$\alpha$ structure sampled by the slit rather than the period of any uniquely defined thread.

A similar distinction applies to the amplitude. FILOS uses \(A_\parallel=15\) Mm as the nominal longitudinal displacement scale parallel to the horizontal thread direction \(\mathbf d_j\). The displacement of each thread is subsequently modulated by the spatial weight \(w_j\) and projected onto the image plane. The recovered amplitude of \(12.08\pm0.86\) Mm instead characterizes the collective projected displacement along the slit. The two quantities therefore describe different stages of the forward-modeling chain and do not constitute a direct amplitude-recovery comparison.

The fitted damping time must also be interpreted as a property of the collective signal. The range of thread curvature radii produces a corresponding distribution of oscillation periods, so the initially coherent motion progressively dephases. The fitted value of $219.32\pm47.85$~min is therefore much shorter than the 15.1~h damping time prescribed for the individual threads. This shorter collective timescale is consistent with additional apparent damping associated with the progressive dephasing of the thread ensemble. Other effects associated with projection, image formation, and tracking of the extended H$\alpha$ absorption feature may also contribute, so the present experiment does not isolate dephasing as the sole cause of the recovered damping.

To further investigate the contribution of dephasing, we repeated the reference experiment with a common oscillation period assigned to all threads, keeping the remaining configuration unchanged (Appendix~\ref{app:dephasing}). The recovered collective damping time increases from $219\pm48$~min to $732\pm132$~min, supporting the interpretation that period dispersion contributes substantially to the apparent damping of the collective motion. A systematic investigation of the relative contributions of dephasing, projection, and image formation using FILOS is beyond the scope of the present work and is left for future studies.

\begin{table}[!ht]
\caption{Relevant FILOS inputs and quantities recovered from the collective time--distance analysis of the reference simulation. The prescribed longitudinal amplitude is the nominal displacement scale before the spatial weighting in Eq.~(\ref{eq:spatial_weight}) and image-plane projection. The fitted quantities characterize the collective projected H$\alpha$ motion sampled by the slit and therefore do not all correspond directly to individual-thread parameters.}
\label{tab:oscillation_fit}
\centering
\small
\begin{tabular}{@{}ll@{}}
\hline\hline
Quantity & Value \\
\hline
Expected thread-period range               & 44.2--87.6 min \\
Median expected thread period              & 60.8 min \\
Recovered collective period                & $57.71 \pm 0.35$ min \\
\hline
Prescribed onset time                      & 20 min \\
Recovered onset time                       & $19.53 \pm 0.23$ min \\
\hline
Prescribed longitudinal amplitude scale    & 15 Mm \\
Recovered projected slit amplitude         & $12.08 \pm 0.86$ Mm \\
\hline
Prescribed thread damping time             & 15.1 h \\
Recovered collective damping time          & $219.32 \pm 47.85$ min \\
\hline
\end{tabular}
\end{table}
\subsection{Derived products from FILOS}

In addition to the synthetic H$\alpha$ observations, FILOS provides direct
access to quantities generated internally by the forward model. These
products retain information that is known exactly from the synthetic
filament and can therefore be used as model-defined reference for subsequent analysis.

\subsubsection{Filament segmentation}
\begin{figure*}[!ht]
    \sidecaption
    \includegraphics[width=12cm]{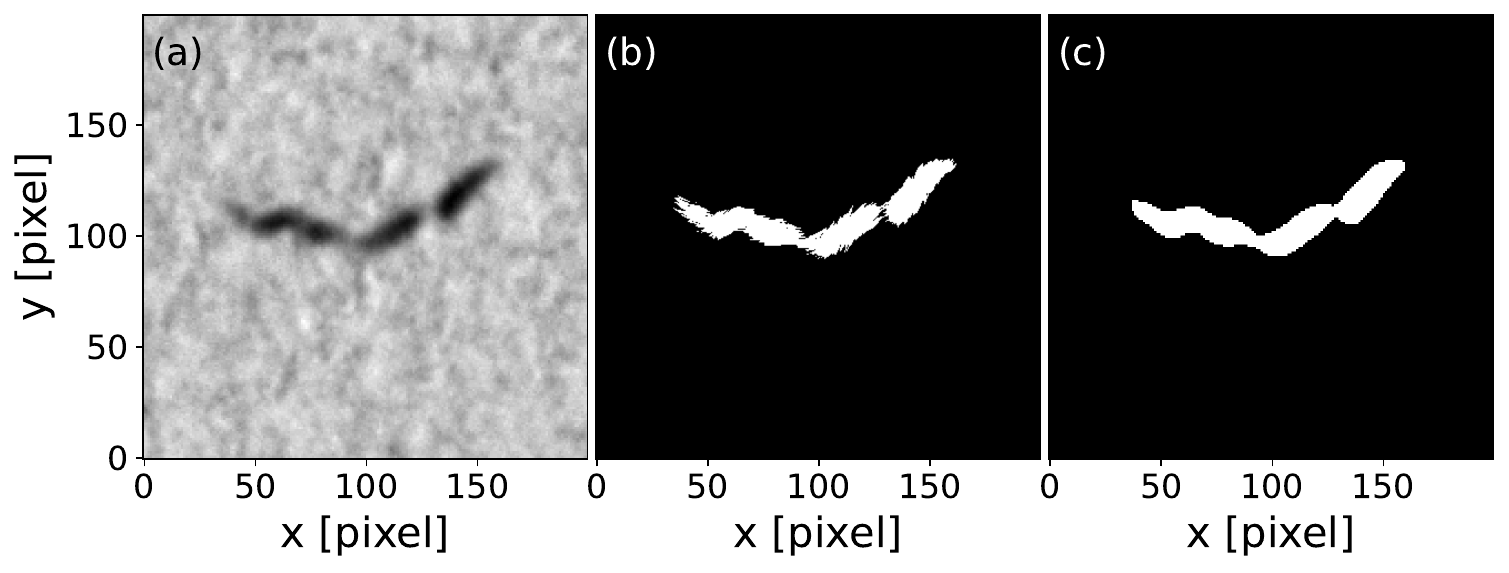}
    \caption{First frame and FILOS-produced segmentation masks from the experiment described in \cref{subsec:longitudinal}. For this experiment we assumed a threshold for the opacity of $\tau_{th} = 0.1$. (a) Rendered H$\alpha$ observation. (b) High-resolution ground-truth mask derived from the filament opacity. (c) Mask degraded to the GONG spatial resolution, providing a ground-truth label suitable for training and validating filament-segmentation models.}
    \label{fig:segmentation}
\end{figure*}

Automatic segmentation of solar filaments in \ha observations remains a challenging problem because of their irregular morphology, variable contrast across the solar disk, fine internal structure, and sensitivity to noise and observational artifacts. Early approaches relied mainly on thresholding, region-growing, and morphological operations \citep{Gao...2002, Shih...2003, Fuller...2004, Bernasconi...2005}, while more recent methods increasingly use deep-learning architectures such as CNNs, U-Nets, and instance-segmentation models \citep{Zhu...2019, Liu...2021, Guo...2022, Diercke...2024, Reche...2026}. The development of these methods depends on reliable pixel-level reference annotations. Public resources such as the MAGFiLO dataset \citep{MagFiLo...2024} and the SWEFil dataset introduced by \citet{Reche...2026} provide manually annotated GONG \ha observations for training and benchmarking segmentation models. However, constructing such datasets requires substantial expert annotation, particularly because filament boundaries can become ambiguous for low-contrast structures, fine-scale features, and regions affected by observational degradation.

FILOS provides a complementary approach for generating such labeled datasets, as illustrated in \cref{fig:segmentation}. Rather than deriving the segmentation from the rendered \ha\ intensity image, the reference mask is constructed directly from the line-center optical-depth distribution of the synthetic filament. At high resolution, a binary mask is obtained by applying a selected optical-depth threshold to the opacity map, thereby tracing the projected absorbing structure before instrumental degradation. FILOS also propagates the opacity information through the instrumental model: the filament transmission is convolved with the point-spread function and integrated over the detector pixels to obtain the effective optical depth at the resolution of the synthetic GONG image. Applying the same threshold to this degraded opacity map produces the corresponding GONG-resolution segmentation mask. Thus, the low-resolution label accounts directly for the loss of spatial information introduced by the instrument rather than being obtained by simply resampling the high-resolution binary mask. For the reference experiment, we adopt \(\tau_{\rm th}=0.1\). 

This procedure is applied independently to every frame of a synthetic sequence, producing temporally resolved image--mask pairs that follow the imposed filament dynamics. FILOS can therefore be used to complement existing observational datasets such as MAGFiLO and SWEFil by generating controlled training and validation samples without additional manual annotation. The simulation parameters can also be varied systematically, providing datasets that span different filament morphologies, optical properties, dynamics, and observational conditions while retaining a model-defined pixel-level reference segmentation.

\subsubsection{Velocity field maps from the dynamics}
FILOS also provides two-dimensional velocity fields derived directly from the prescribed filament dynamics. At each time step, the longitudinal and transverse velocities of every thread are evaluated analytically from the same damped oscillation model used to evolve its position. These components are combined according to the orientation of each thread and projected onto the image plane, accounting for the position of the filament on the solar disk.  

Because several threads can contribute to the same image pixel, FILOS constructs the local velocity as an opacity-weighted average of their projected velocities. For each image-plane component $c\in\{x,y\}$, this can be expressed as
\begin{equation}
    v_c(x,y,t) =
    \frac{\sum_j \tau_j(x,y,t)\,v_{j,c}(t)}
         {\sum_j \tau_j(x,y,t)},
\end{equation}
where $\tau_j(x,y,t)$ is the optical-depth contribution of thread $j$ and $v_{j,c}(t)$ is its projected velocity. The numerator and optical-depth map are subsequently convolved with the instrumental PSF and degraded to the GONG spatial resolution before computing the final velocity field. Thus, regions where several threads overlap naturally represent the collective motion of the material contributing most strongly to the H$\alpha$ absorption. 

These maps should be interpreted as opacity-weighted material velocities of the synthetic thread ensemble. Apparent feature velocities inferred from image-tracking or optical-flow methods describe the motion of intensity structures and may therefore differ from these material-velocity reference fields.

\begin{figure*}
    \centering
    \includegraphics[width=\linewidth]{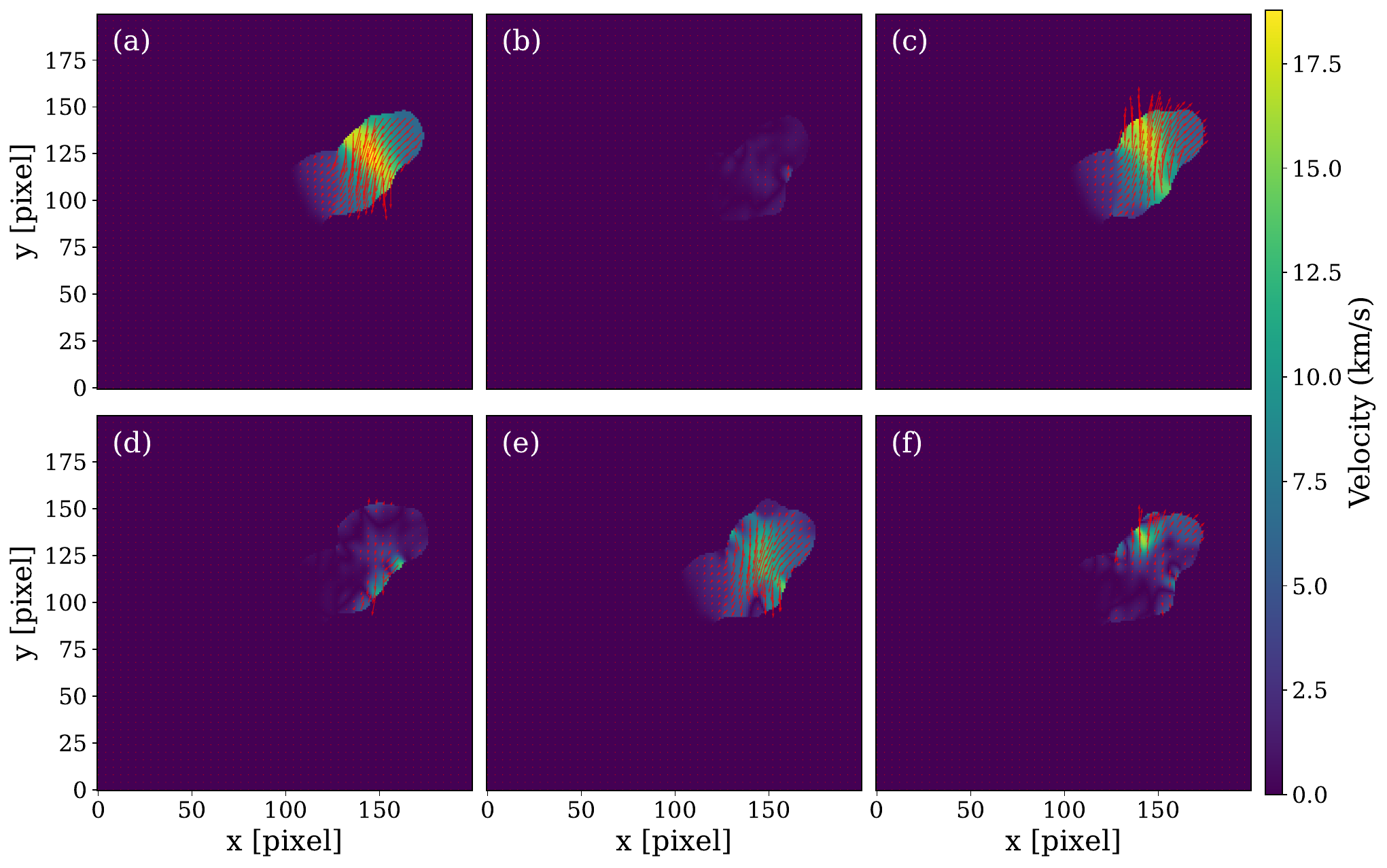}
    \caption{Two-dimensional velocity field derived from the FILOS dynamics for the same frames shown in \cref{fig:video_frames}. Panels (a)--(f) are separated by 15 min. Red arrows indicate the direction of the projected velocity field, while the background colormap shows the velocity magnitude in km\,s$^{-1}$. The associated movie is available online.}
    \label{fig:velocity_maps}
\end{figure*}

Figure~\ref{fig:velocity_maps} shows representative velocity maps from the reference simulation. In addition to the synthetic H$\alpha$ observations, these maps provide the underlying plane-of-sky motion associated with the imposed dynamics and therefore constitute an additional model-defined reference product that can be used to develop and test methods for recovering filament motions from observations.

\section{Discussion and conclusions}
In this work, we presented FILOS, a forward model for generating controlled synthetic H$\alpha$ observations of oscillating solar filaments. The model combines the observed morphology of real filaments with a prescribed three-dimensional ensemble of threads, a static thermodynamic structure, an approximate non-LTE treatment of the H$\alpha$ opacity, and controlled kinematic motions. The resulting filament is projected according to its position on the solar disk, degraded to the spatial resolution of GONG, and rendered over real GONG H$\alpha$ observations. FILOS therefore produces synthetic sequences that retain the complexity of observed chromospheric backgrounds while providing model-defined reference labels for the geometry and dynamics of the inserted filament.

This combination of observational realism and controlled dynamics is the main purpose of the model. In observational data, the physical parameters of an oscillation must be inferred from the evolution of the observed intensity, and the accuracy of this inference is generally unknown. With FILOS, the prescribed inputs and the intermediate quantities generated by the forward model are known, allowing analysis outputs to be compared with the appropriate model-defined reference. The model can therefore be used both to study how projection, spatial resolution, background structure, filament morphology, and internal thread properties affect the observable signatures of an oscillation and to generate labeled datasets for training and evaluating motion-inference and segmentation networks.

We illustrated these capabilities with a representative longitudinal oscillation. The synthetic filament was constructed from 5392 threads distributed along a measured 154.1~Mm spine. The longitudinal dynamics were assigned using the pendulum model from \citet{LunaModel...2012, Luna...2022}, producing thread periods between 44.2 and 87.6~min, with a median of 60.8~min. The resulting GONG-like sequence shows the expected displacement of the absorbing filament material and reproduces the characteristic signatures of this motion in both the spatial and frequency domains. In particular, the standard-normalized Lomb–Scargle power maps concentrate the oscillatory signal around the dominant period range and reproduce the period-halving signature at the equilibrium position that arises when the oscillating absorbing structure repeatedly crosses the same image location.

A conventional time--distance analysis of this experiment illustrates how thread-level dynamics are transformed into collective observables. The recovered onset time, $19.53\pm0.23$~min, closely matches the prescribed value of 20~min. The fitted period, $57.71\pm0.35$~min, lies within the imposed thread-period distribution and is close to its median of 60.8~min. This fitted period characterizes the collective H$\alpha$ displacement sampled by the slit rather than the motion of any uniquely defined thread. Similarly, the recovered amplitude of $12.08\pm0.86$~Mm represents the projected image-plane displacement along the slit, whereas FILOS prescribes a nominal 15 Mm horizontal displacement along each thread's direction, before spatial weighting and projection onto the image plane. These quantities therefore probe different stages of the forward-modeling chain rather than constituting a direct parameter-by-parameter recovery test.

The damping behavior further illustrates this distinction between prescribed thread dynamics and the observed collective motion. Because the curvature radius varies across the filament, the individual threads oscillate with different periods and progressively move out of phase. The H$\alpha$ absorption pattern therefore loses spatial coherence with time, producing an apparent decay of the bulk displacement in the time--distance diagram in addition to the damping prescribed for each thread. The progressive dephasing of the thread ensemble provides an additional source of apparent damping of the collective H$\alpha$ displacement and is consistent with the substantially shorter damping time recovered from the time--distance diagram. The present experiment does not separately quantify the contributions of dephasing, projection, image formation, and trajectory tracking to this fitted timescale.

In addition to the rendered H$\alpha$ sequences, FILOS provides access to quantities that are defined by construction in the forward model. The opacity distribution produces high-resolution filament masks that can subsequently be degraded to the same resolution as the synthetic observations. This generates image--mask pairs with model-defined reference masks, providing datasets for training and evaluating filament segmentation methods. Similarly, the prescribed dynamics provide opacity-weighted plane-of-sky velocity fields at each time step. These maps provide velocity labels for supervised methods designed to estimate filament motions from imaging observations.

These products extend the application of FILOS beyond the analysis of an individual synthetic oscillation. Large ensembles can be generated while systematically varying filament geometry, oscillation amplitude, period, damping time, excitation location, spatial coherence, viewing angle, and observational conditions. Such experiments can support the training and evaluation of automatic inference methods and quantify detection efficiency and parameter-recovery biases across a known parameter space. This is particularly relevant for automatic methods developed to analyze large GONG datasets, including the spectral detection framework presented in \citet{Castello...2026}. Synthetic observations provide cases for which the presence, location, and properties of an oscillation are known independently of the detection method itself, allowing completeness and sensitivity to be studied under controlled conditions.

Beyond oscillation analysis, the same framework could eventually be extended to generate labeled examples of more general filament motions. Because FILOS provides synthetic H$\alpha$ sequences together with model-defined plane-of-sky velocity fields, it could support the development of methods for general motion inference and long-term filament tracking. Such an extension would require slowly evolving filament morphologies, longer temporal sequences, and validation against observations. Whether these data could also help identify changes preceding an eruption is an interesting longer-term question, but it lies beyond the scope of the present work.

The current implementation adopts a deliberately simplified physical description in order to preserve this level of control. The filament consists of circularly dipped threads with a prescribed static thermodynamic structure, while the temporal evolution is introduced kinematically and the corresponding opacity distribution is transported with the threads. H$\alpha$ image formation is represented through an approximate opacity calculation, a common source function, and an effective treatment of the GONG spectral response. In particular, the present effective GONG spectral response does not include velocity-dependent visibility changes associated with Doppler shifts through the instrumental passband. These assumptions make it possible to generate large numbers of synthetic observations efficiently while retaining the main geometrical and observational properties required for the applications considered here. Future extensions can introduce more general thread geometries, additional forms of filament motion, and more detailed treatments of the temporal plasma and radiative evolution when required by a specific application.

FILOS establishes a framework for connecting controlled models of filament dynamics with the measurements obtained from H$\alpha$ imaging. Rather than having access only to the final observable, the complete synthetic filament and its evolution are available throughout the forward-modeling process. This makes it possible to determine how known physical and geometrical inputs are transformed into observable signatures and how accurately those inputs can subsequently be recovered. We therefore expect FILOS to support the development and validation of automatic oscillation-analysis and motion-inference methods, the generation of labeled datasets for image segmentation, and systematic studies of the observational biases affecting the characterization of solar filament oscillations.

\section*{Data availability}
The FILOS repository can be found in \url{https://github.com/GuillemCastello/FILOS}, with a complete description, tutorial and examples.

\begin{acknowledgements}
This publication is part of the I+D+i project PID2023-147708NB-I00 funded by MICIU/AEI/10.13039/501100011033 and by  FEDER, EU. We thank Stanislav Gunár for his constructive comments on the manuscript. G. Castelló acknowledges financial support from the Direcció General d'Universitats, Recerca i Ensenyaments Artístics Superiors of the Government of the Balearic Islands through a pre-doctoral fellowship co-financed by the European Social Fund Plus (FSE+) within the framework of the Balearic Islands Programme 2021–2027. Co-funded by the European Union.
\end{acknowledgements}

\bibliographystyle{aa.bst}
\bibliography{bibliography.bib}

@ARTICLE{Gunar...2013,
       author = {{Gun{\'a}r}, S. and {Mackay}, D.~H. and {Anzer}, U. and {Heinzel}, P.},
        title = "{Non-linear force-free magnetic dip models of quiescent prominence fine structures}",
      journal = {\aap},
         year = 2013,
        month = mar,
       volume = {551},
          eid = {A3},
        pages = {A3},
          doi = {10.1051/0004-6361/201220597},
       adsurl = {https://ui.adsabs.harvard.edu/abs/2013A&A...551A...3G}
}

@ARTICLE{Asensio...2017,
       author = {{Asensio Ramos}, A. and {Requerey}, I.~S. and {Vitas}, N.},
        title = "{DeepVel: Deep learning for the estimation of horizontal velocities at the solar surface}",
      journal = {\aap},
         year = 2017,
        volume = {604},
          eid = {A11},
        pages = {A11},
          doi = {10.1051/0004-6361/201730783}
}

@ARTICLE{Liu...2025,
       author = {{Liu}, Jiajia and {Xie}, Quan and {Zhong}, Cheng and {Ye}, Yudong and {Wang}, Yimin and {Ji}, Kaifan and {Wang}, Yuming},
        title = "{Tracking the photospheric horizontal velocity field with shallow U-net models}",
      journal = {\aap},
         year = 2025,
        volume = {698},
          eid = {A263},
        pages = {A263},
          doi = {10.1051/0004-6361/202453627}
}

@ARTICLE{Gunar...2015,
       author = {{Gun{\'a}r}, Stanislav and {Mackay}, Duncan H.},
        title = "{3D Whole-Prominence Fine Structure Modeling}",
      journal = {\apj},
         year = 2015,
        month = apr,
       volume = {803},
       number = {2},
          eid = {64},
        pages = {64},
          doi = {10.1088/0004-637X/803/2/64},
       adsurl = {https://ui.adsabs.harvard.edu/abs/2015ApJ...803...64G}
}

@ARTICLE{GunarII...2015,
       author = {{Gun{\'a}r}, Stanislav and {Mackay}, Duncan H.},
        title = "{3D Whole-prominence Fine Structure Modeling. II. Prominence Evolution}",
      journal = {\apj},
         year = 2015,
        month = oct,
       volume = {812},
       number = {2},
          eid = {93},
        pages = {93},
          doi = {10.1088/0004-637X/812/2/93},
       adsurl = {https://ui.adsabs.harvard.edu/abs/2015ApJ...812...93G}
}

@ARTICLE{Heinzel...2015,
       author = {{Heinzel}, P. and {Gun{\'a}r}, S. and {Anzer}, U.},
        title = "{Fast approximate radiative transfer method for visualizing the fine structure of prominences in the hydrogen H{\ensuremath{\alpha}} line}",
      journal = {\aap},
         year = 2015,
        month = jul,
       volume = {579},
          eid = {A16},
        pages = {A16},
          doi = {10.1051/0004-6361/201525716},
       adsurl = {https://ui.adsabs.harvard.edu/abs/2015A&A...579A..16H}
}

@ARTICLE{Gunar...2018,
       author = {{Gun{\'a}r}, Stanislav and {Dud{\'\i}k}, Jaroslav and {Aulanier}, Guillaume and {Schmieder}, Brigitte and {Heinzel}, Petr},
        title = "{Importance of the H{\ensuremath{\alpha}} Visibility and Projection Effects for the Interpretation of Prominence Fine-structure Observations}",
      journal = {\apj},
         year = 2018,
        month = nov,
       volume = {867},
       number = {2},
          eid = {115},
        pages = {115},
          doi = {10.3847/1538-4357/aae4e1},
       adsurl = {https://ui.adsabs.harvard.edu/abs/2018ApJ...867..115G}
}

@ARTICLE{Osborne...2021,
       author = {{Osborne}, Christopher M.~J. and {Mili{\'c}}, Ivan},
        title = "{The Lightweaver Framework for Nonlocal Thermal Equilibrium Radiative Transfer in Python}",
      journal = {\apj},
         year = 2021,
        month = aug,
       volume = {917},
       number = {1},
          eid = {14},
        pages = {14},
          doi = {10.3847/1538-4357/ac02be},
archivePrefix = {arXiv},
       eprint = {2107.00475},
 primaryClass = {astro-ph.IM},
       adsurl = {https://ui.adsabs.harvard.edu/abs/2021ApJ...917...14O}
}

@MISC{Osborne...2022,
       author = {{Osborne}, Christopher M.~J.},
        title = "{Goobley/Promweaver: v0.1.1}",
         year = 2022,
        month = may,
          doi = {10.5281/zenodo.6546897},
    publisher = {Zenodo}
}

@ARTICLE{Jenkins...2023,
       author = {{Jenkins}, J.~M. and {Osborne}, C.~M.~J. and {Keppens}, R.},
        title = "{1.5D non-LTE spectral synthesis of a 3D filament and prominence simulation}",
      journal = {\aap},
         year = 2023,
        month = feb,
       volume = {670},
          eid = {A179},
        pages = {A179},
          doi = {10.1051/0004-6361/202244868},
archivePrefix = {arXiv},
       eprint = {2211.14869},
 primaryClass = {astro-ph.SR},
       adsurl = {https://ui.adsabs.harvard.edu/abs/2023A&A...670A.179J}
}

@ARTICLE{Jenkins...2024,
       author = {{Jenkins}, Jack M. and {Osborne}, Christopher M.~J. and {Qiu}, Ye and {Keppens}, Rony and {Li}, Chuan},
        title = "{The Bright Rim Prominences according to 2.5D Radiative Transfer}",
      journal = {\apjl},
         year = 2024,
        month = apr,
       volume = {964},
       number = {2},
          eid = {L34},
        pages = {L34},
          doi = {10.3847/2041-8213/ad3423},
archivePrefix = {arXiv},
       eprint = {2403.09931},
 primaryClass = {astro-ph.SR},
       adsurl = {https://ui.adsabs.harvard.edu/abs/2024ApJ...964L..34J}
}

@ARTICLE{Luna...2012,
       author = {{Luna}, M. and {Karpen}, J.},
        title = "{Large-amplitude Longitudinal Oscillations in a Solar Filament}",
      journal = {\apjl},
         year = 2012,
        month = may,
       volume = {750},
       number = {1},
          eid = {L1},
        pages = {L1},
          doi = {10.1088/2041-8205/750/1/L1},
archivePrefix = {arXiv},
       eprint = {1203.5027},
 primaryClass = {astro-ph.SR},
       adsurl = {https://ui.adsabs.harvard.edu/abs/2012ApJ...750L...1L}
}

@ARTICLE{LunaModel...2012,
       author = {{Luna}, M. and {D{\'\i}az}, A.~J. and {Karpen}, J.},
        title = "{The Effects of Magnetic-field Geometry on Longitudinal Oscillations of Solar Prominences}",
      journal = {\apj},
         year = 2012,
        month = sep,
       volume = {757},
       number = {1},
          eid = {98},
        pages = {98},
          doi = {10.1088/0004-637X/757/1/98},
archivePrefix = {arXiv},
       eprint = {1207.6358},
 primaryClass = {astro-ph.SR},
       adsurl = {https://ui.adsabs.harvard.edu/abs/2012ApJ...757...98L}
}

@ARTICLE{Luna...2016,
       author = {{Luna}, M. and {D{\'\i}az}, A.~J. and {Oliver}, R. and {Terradas}, J. and {Karpen}, J.},
        title = "{The effects of magnetic-field geometry on longitudinal oscillations of solar prominences: Cross-sectional area variation for thin tubes}",
      journal = {\aap},
         year = 2016,
        month = sep,
       volume = {593},
          eid = {A64},
        pages = {A64},
          doi = {10.1051/0004-6361/201628845},
archivePrefix = {arXiv},
       eprint = {1607.02996},
 primaryClass = {astro-ph.SR},
       adsurl = {https://ui.adsabs.harvard.edu/abs/2016A&A...593A..64L}
}

@ARTICLE{Luna...2022,
       author = {{Luna}, M. and {Terradas}, J. and {Karpen}, J. and {Ballester}, J.~L.},
        title = "{Extension and validation of the pendulum model for longitudinal solar prominence oscillations}",
      journal = {\aap},
         year = 2022,
        month = apr,
       volume = {660},
          eid = {A54},
        pages = {A54},
          doi = {10.1051/0004-6361/202142907},
archivePrefix = {arXiv},
       eprint = {2202.07957},
 primaryClass = {astro-ph.SR},
       adsurl = {https://ui.adsabs.harvard.edu/abs/2022A&A...660A..54L}
}

@ARTICLE{LunaSpectral...2022,
       author = {{Luna}, M. and {M{\'e}rou Mestre}, J.~R. and {Auch{\`e}re}, F.},
        title = "{Automatic detection technique for solar filament oscillations in GONG data}",
      journal = {\aap},
         year = 2022,
        month = oct,
       volume = {666},
          eid = {A195},
        pages = {A195},
          doi = {10.1051/0004-6361/202244181},
archivePrefix = {arXiv},
       eprint = {2209.05087},
 primaryClass = {astro-ph.SR},
       adsurl = {https://ui.adsabs.harvard.edu/abs/2022A&A...666A.195L}
}

@ARTICLE{LunaCatalog...2018,
       author = {{Luna}, M. and {Karpen}, J. and {Ballester}, J.~L. and {Muglach}, K. and {Terradas}, J. and {Kucera}, T. and {Gilbert}, H.},
        title = "{GONG Catalog of Solar Filament Oscillations Near Solar Maximum}",
      journal = {\apjs},
         year = 2018,
        month = jun,
       volume = {236},
       number = {2},
          eid = {35},
        pages = {35},
          doi = {10.3847/1538-4365/aabde7},
archivePrefix = {arXiv},
       eprint = {1804.03743},
 primaryClass = {astro-ph.SR},
       adsurl = {https://ui.adsabs.harvard.edu/abs/2018ApJS..236...35L}
}

@ARTICLE{LabrosseReview...2010,
       author = {{Labrosse}, N. and {Heinzel}, P. and {Vial}, J.-C. and {Kucera}, T. and {Parenti}, S. and {Gun{\'a}r}, S. and {Schmieder}, B. and {Kilper}, G.},
        title = "{Physics of Solar Prominences: I{\textemdash}Spectral Diagnostics and Non-LTE Modelling}",
      journal = {\ssr},
         year = 2010,
        month = apr,
       volume = {151},
       number = {4},
        pages = {243-332},
          doi = {10.1007/s11214-010-9630-6},
archivePrefix = {arXiv},
       eprint = {1001.1620},
 primaryClass = {astro-ph.SR},
       adsurl = {https://ui.adsabs.harvard.edu/abs/2010SSRv..151..243L}
}

@ARTICLE{HeinzelReview...2025,
       author = {{Heinzel}, Petr and {Gun{\'a}r}, Stanislav},
        title = "{Radiative Transfer in Solar Prominences: An Overview and Current Trends}",
      journal = {\solphys},
         year = 2025,
        month = nov,
       volume = {300},
       number = {12},
          eid = {167},
        pages = {167},
          doi = {10.1007/s11207-025-02585-y},
       adsurl = {https://ui.adsabs.harvard.edu/abs/2025SoPh..300..167H}
}

@ARTICLE{GibsonReview...2018,
       author = {{Gibson}, Sarah E.},
        title = "{Solar prominences: theory and models. Fleshing out the magnetic skeleton}",
      journal = {Living Reviews in Solar Physics},
         year = 2018,
        month = dec,
       volume = {15},
       number = {1},
          eid = {7},
        pages = {7},
          doi = {10.1007/s41116-018-0016-2},
       adsurl = {https://ui.adsabs.harvard.edu/abs/2018LRSP...15....7G}
}

@ARTICLE{Castello...2025,
       author = {{Castell{\'o}}, G. and {Luna}, M. and {Terradas}, J.},
        title = "{Fast Bayesian spectral analysis using convolutional neural networks: Applications to GONG H{\ensuremath{\alpha}} solar data}",
      journal = {\aap},
         year = 2025,
        month = feb,
       volume = {694},
          eid = {A237},
        pages = {A237},
          doi = {10.1051/0004-6361/202452928},
archivePrefix = {arXiv},
       eprint = {2501.12743},
 primaryClass = {astro-ph.SR},
       adsurl = {https://ui.adsabs.harvard.edu/abs/2025A&A...694A.237C}
}

@ARTICLE{Castello...2026,
       author = {{Castell{\'o}}, G. and {Luna}, M. and {Terradas}, J.},
        title = "{Automatic detection of solar filament oscillations: I. Multi-scale spectral pipeline}",
      journal = {\aap},
         year = 2026,
        month = aug,
       volume = {712},
          eid = {A44},
        pages = {A44},
          doi = {10.1051/0004-6361/202660956},
       adsurl = {https://ui.adsabs.harvard.edu/abs/2026A&A...712A..44C}
}

@ARTICLE{Arregui...2018,
	title = {Prominence oscillations},
	volume = {15},
	url = {http://adsabs.harvard.edu/cgi-bin/nph-data_query?bibcode=2018LRSP...15....3A&link_type=EJOURNAL},
	doi = {10.1007/s41116-018-0012-6},
	number = {1},
	journal = {Living Rev. Sol. Phys.},
	author = {Arregui, Iñigo and Oliver, Ramón and Ballester, Jose Luis},
	year = {2018},
	note = {Publisher: Springer International Publishing},
	pages = {3},
}

@ARTICLE{Mackay...2010,
       author = {{Mackay}, D.~H. and {Karpen}, J.~T. and {Ballester}, J.~L. and {Schmieder}, B. and {Aulanier}, G.},
        title = "{Physics of Solar Prominences: II{\textemdash}Magnetic Structure and Dynamics}",
      journal = {\ssr},
         year = 2010,
        month = apr,
       volume = {151},
       number = {4},
        pages = {333-399},
          doi = {10.1007/s11214-010-9628-0},
archivePrefix = {arXiv},
       eprint = {1001.1635},
 primaryClass = {astro-ph.SR},
       adsurl = {https://ui.adsabs.harvard.edu/abs/2010SSRv..151..333M}
}

@ARTICLE{Oliver...2002,
       author = {{Oliver}, Ram{\'o}n and {Ballester}, Jos{\'e} Luis},
        title = "{Oscillations in Quiescent Solar Prominences Observations and Theory   (Invited Review)}",
      journal = {\solphys},
         year = 2002,
        month = mar,
       volume = {206},
       number = {1},
        pages = {45-67},
          doi = {10.1023/A:1014915428440},
       adsurl = {https://ui.adsabs.harvard.edu/abs/2002SoPh..206...45O}
}

@ARTICLE{Zhang...2012,
	title = {Observations and simulations of longitudinal oscillations of an active region prominence},
	volume = {542},
	copyright = {© ESO, 2012},
	issn = {0004-6361, 1432-0746},
	url = {https://www.aanda.org/articles/aa/abs/2012/06/aa18786-12/aa18786-12.html},
	doi = {10.1051/0004-6361/201218786},
	language = {en},
	urldate = {2020-10-19},
	journal = {\aap},
	publisher = {EDP Sciences},
	author = {Zhang, Q. M. and Chen, P. F. and Xia, C. and Keppens, R.},
	month = jun,
	year = {2012},
	pages = {A52},
}

@ARTICLE{Diercke...2024,
       author = {{Diercke}, Andrea and {Jarolim}, Robert and {Kuckein}, Christoph and {Gonz{\'a}lez Manrique}, Sergio J. and {Ziener}, Marco and {Veronig}, Astrid M. and {Denker}, Carsten and {P{\"o}tzi}, Werner and {Podladchikova}, Tatiana and {Pevtsov}, Alexei A.},
        title = "{A universal method for solar filament detection from H{\ensuremath{\alpha}} observations using semi-supervised deep learning}",
      journal = {\aap},
         year = {2024},
        month = {Jun},
       volume = {686},
          eid = {A213},
        pages = {A213},
          doi = {10.1051/0004-6361/202348314},
archivePrefix = {arXiv},
       eprint = {2402.15407},
 primaryClass = {astro-ph.SR},
       adsurl = {https://ui.adsabs.harvard.edu/abs/2024A&A...686A.213D}
}

@ARTICLE{Reche...2026,
       author = {{Reche}, Antonio and {Cid}, Consuelo},
        title = "{New dataset and framework for filament detection, classification, and segmentation using deep learning}",
      journal = {Acta Astronaut.},
         year = 2026,
        month = dec,
       volume = {249},
        pages = {381-398},
          doi = {10.1016/j.actaastro.2026.07.010},
       adsurl = {https://ui.adsabs.harvard.edu/abs/2026AcAau.249..381R}
}

@ARTICLE{Gao...2002,
       author = {{Gao}, Jianlin and {Wang}, Haimin and {Zhou}, Mengchu},
        title = "{Development of an Automatic Filament Disappearance Detection System}",
      journal = {\solphys},
         year = {2002},
        month = {Jan},
       volume = {205},
       number = {1},
        pages = {93-103},
          doi = {10.1023/A:1013851808367},
       adsurl = {https://ui.adsabs.harvard.edu/abs/2002SoPh..205...93G}
}

@ARTICLE{Shih...2003,
       author = {{Shih}, Frank Y. and {Kowalski}, Artur J.},
        title = "{Automatic Extraction of Filaments in H{\ensuremath{\alpha}} Solar Images}",
      journal = {\solphys},
         year = {2003},
        month = {Dec},
       volume = {218},
       number = {1},
        pages = {99-122},
          doi = {10.1023/B:SOLA.0000013052.34180.58},
       adsurl = {https://ui.adsabs.harvard.edu/abs/2003SoPh..218...99S}
}

@InProceedings{Fuller...2004,
    author="Fuller, N.
    and Aboudarham, J.",
    editor="Negoita, Mircea Gh.
    and Howlett, Robert J.
    and Jain, Lakhmi C.",
    title="Automatic Detection of Solar Filaments Versus Manual Digitization",
    booktitle="Knowledge-Based Intelligent Information and Engineering Systems",
    year="2004",
    publisher="Springer Berlin Heidelberg",
    address="Berlin, Heidelberg",
    pages="467--475",
    isbn="978-3-540-30134-9"
}

@ARTICLE{Bernasconi...2005,
       author = {{Bernasconi}, Pietro N. and {Rust}, David M. and {Hakim}, Daniel},
        title = "{Advanced Automated Solar Filament Detection And Characterization Code: Description, Performance, And Results}",
      journal = {\solphys},
         year = {2005},
        month = {May},
       volume = {228},
       number = {1-2},
        pages = {97-117},
          doi = {10.1007/s11207-005-2766-y},
       adsurl = {https://ui.adsabs.harvard.edu/abs/2005SoPh..228...97B}
}

@ARTICLE{Zhu...2019,
       author = {{Zhu}, Gaofei and {Lin}, Ganghua and {Wang}, Dongguang and {Liu}, Suo and {Yang}, Xiao},
        title = "{Solar Filament Recognition Based on Deep Learning}",
      journal = {\solphys},
         year = {2019},
        month = {Sep},
       volume = {294},
       number = {9},
          eid = {117},
        pages = {117},
          doi = {10.1007/s11207-019-1517-4},
archivePrefix = {arXiv},
       eprint = {1909.06580},
 primaryClass = {astro-ph.IM},
       adsurl = {https://ui.adsabs.harvard.edu/abs/2019SoPh..294..117Z}
}

@ARTICLE{Liu...2021,
       author = {{Liu}, Dan and {Song}, Wei and {Lin}, Ganghua and {Wang}, Haimin},
        title = "{Solar Filament Segmentation Based on Improved U-Nets}",
      journal = {\solphys},
         year = {2021},
        month = {Dec},
       volume = {296},
       number = {12},
          eid = {176},
        pages = {176},
          doi = {10.1007/s11207-021-01920-3},
       adsurl = {https://ui.adsabs.harvard.edu/abs/2021SoPh..296..176L}
}

@ARTICLE{Guo...2022,
       author = {{Guo}, Xulong and {Yang}, Yunfei and {Feng}, Song and {Bai}, Xianyong and {Liang}, Bo and {Dai}, Wei},
        title = "{Solar-Filament Detection and Classification Based on Deep Learning}",
      journal = {\solphys},
         year = {2022},
        month = {Aug},
       volume = {297},
       number = {8},
          eid = {104},
        pages = {104},
          doi = {10.1007/s11207-022-02019-z},
       adsurl = {https://ui.adsabs.harvard.edu/abs/2022SoPh..297..104G}
}

@article{MagFiLo...2024,
	title = {A dataset of manually annotated filaments from {H}-alpha observations},
	volume = {11},
	issn = {2052-4463},
	url = {https://doi.org/10.1038/s41597-024-03876-y},
	doi = {10.1038/s41597-024-03876-y},
	number = {1},
	journal = {Scientific Data},
	author = {Ahmadzadeh, Azim and Adhyapak, Rohan and Chaurasiya, Kartik and Nagubandi, Laxmi Alekhya and Aparna, V. and Martens, Petrus C. and Pevtsov, Alexei and Bertello, Luca and Pevtsov, Alexander and Douglas, Naomi and McDonald, Samuel and Bawa, Apaar and Kang, Eugene and Wu, Riley and Kempton, Dustin J. and Abdelkarem, Aya and Copeland, Patrick M. and Seelamneni, Sri Harsha},
	month = sep,
	year = {2024},
	pages = {1031},
}

@ARTICLE{Liakh...2020,
       author = {{Liakh}, V. and {Luna}, M. and {Khomenko}, E.},
        title = "{Numerical simulations of large-amplitude oscillations in flux rope solar prominences}",
      journal = {\aap},
         year = 2020,
        month = may,
       volume = {637},
          eid = {A75},
        pages = {A75},
          doi = {10.1051/0004-6361/201937083},
archivePrefix = {arXiv},
       eprint = {2003.04343},
 primaryClass = {astro-ph.SR},
       adsurl = {https://ui.adsabs.harvard.edu/abs/2020A&A...637A..75L}
}

@ARTICLE{Liakh...2021,
       author = {{Liakh}, V. and {Luna}, M. and {Khomenko}, E.},
        title = "{Large-amplitude longitudinal oscillations in solar prominences simulated with different resolutions}",
      journal = {\aap},
         year = 2021,
        month = oct,
       volume = {654},
          eid = {A145},
        pages = {A145},
          doi = {10.1051/0004-6361/202141524},
archivePrefix = {arXiv},
       eprint = {2108.01143},
 primaryClass = {astro-ph.SR},
       adsurl = {https://ui.adsabs.harvard.edu/abs/2021A&A...654A.145L}
}

@ARTICLE{Pietrow...2024,
       author = {{Pietrow}, A.~G.~M. and {Liakh}, V. and {Osborne}, C.~M.~J. and {Jenkins}, J. and {Keppens}, R.},
        title = "{Spectral characteristics of a rotating solar prominence in multiple wavelengths}",
      journal = {\aap},
         year = 2024,
        month = oct,
       volume = {690},
          eid = {L15},
        pages = {L15},
          doi = {10.1051/0004-6361/202452276},
archivePrefix = {arXiv},
       eprint = {2410.03479},
 primaryClass = {astro-ph.SR},
       adsurl = {https://ui.adsabs.harvard.edu/abs/2024A&A...690L..15P}
}

@ARTICLE{ShenWinking...2014,
       author = {{Shen}, Yuandeng and {Liu}, Ying D. and {Chen}, P.~F. and {Ichimoto}, Kiyoshi},
        title = "{Simultaneous Transverse Oscillations of a Prominence and a Filament and Longitudinal Oscillation of Another Filament Induced by a Single Shock Wave}",
      journal = {\apj},
         year = {2014b},
        month = nov,
       volume = {795},
       number = {2},
          eid = {130},
        pages = {130},
          doi = {10.1088/0004-637X/795/2/130},
archivePrefix = {arXiv},
       eprint = {1409.1304},
 primaryClass = {astro-ph.SR},
       adsurl = {https://ui.adsabs.harvard.edu/abs/2014ApJ...795..130S}
}

@article{ShenChain...2014,
doi = {10.1088/0004-637X/786/2/151},
url = {https://doi.org/10.1088/0004-637X/786/2/151},
year = {2014a},
month = {apr},
publisher = {The American Astronomical Society},
volume = {786},
number = {2},
pages = {151},
author = {Shen, Yuandeng and Ichimoto, Kiyoshi and Ishii, Takako T. and Tian, Zhanjun and Zhao, Ruijuan and Shibata, Kazunari},
title = {A CHAIN OF WINKING (OSCILLATING) FILAMENTS TRIGGERED BY AN INVISIBLE EXTREME-ULTRAVIOLET WAVE},
journal = {\apj}
}

@ARTICLE{Tan...2023,
       author = {{Tan}, Song and {Shen}, Yuandeng and {Zhou}, Xinping and {Tang}, Zehao and {Zhou}, Chengrui and {Duan}, Yadan and {Yao}, Surui},
        title = "{Stereoscopic observation of simultaneous longitudinal and transverse oscillations in a single filament driven by two-sided-loop jet}",
      journal = {\mnras},
         year = 2023,
        month = apr,
       volume = {520},
       number = {2},
        pages = {3080-3088},
          doi = {10.1093/mnras/stad295},
archivePrefix = {arXiv},
       eprint = {2301.07989},
 primaryClass = {astro-ph.SR},
       adsurl = {https://ui.adsabs.harvard.edu/abs/2023MNRAS.520.3080T}
}

\begin{appendix}
\nolinenumbers
\section{Effect of thread-period dispersion on collective damping}
\label{app:dephasing}

In Sect.~\ref{subsec:longitudinal}, the reference simulation produces a collective damping time considerably shorter than the damping time prescribed for the individual threads. We attributed part of this difference to progressive dephasing, since the thread periods vary according to their curvature radii and heights. To investigate this effect, we repeated the reference experiment with an otherwise identical configuration, but assigned a common oscillation period of 60 minutes to all threads. The prescribed damping time remained 15.1~h.

\begin{figure}
    \centering
    \includegraphics[width=\columnwidth]{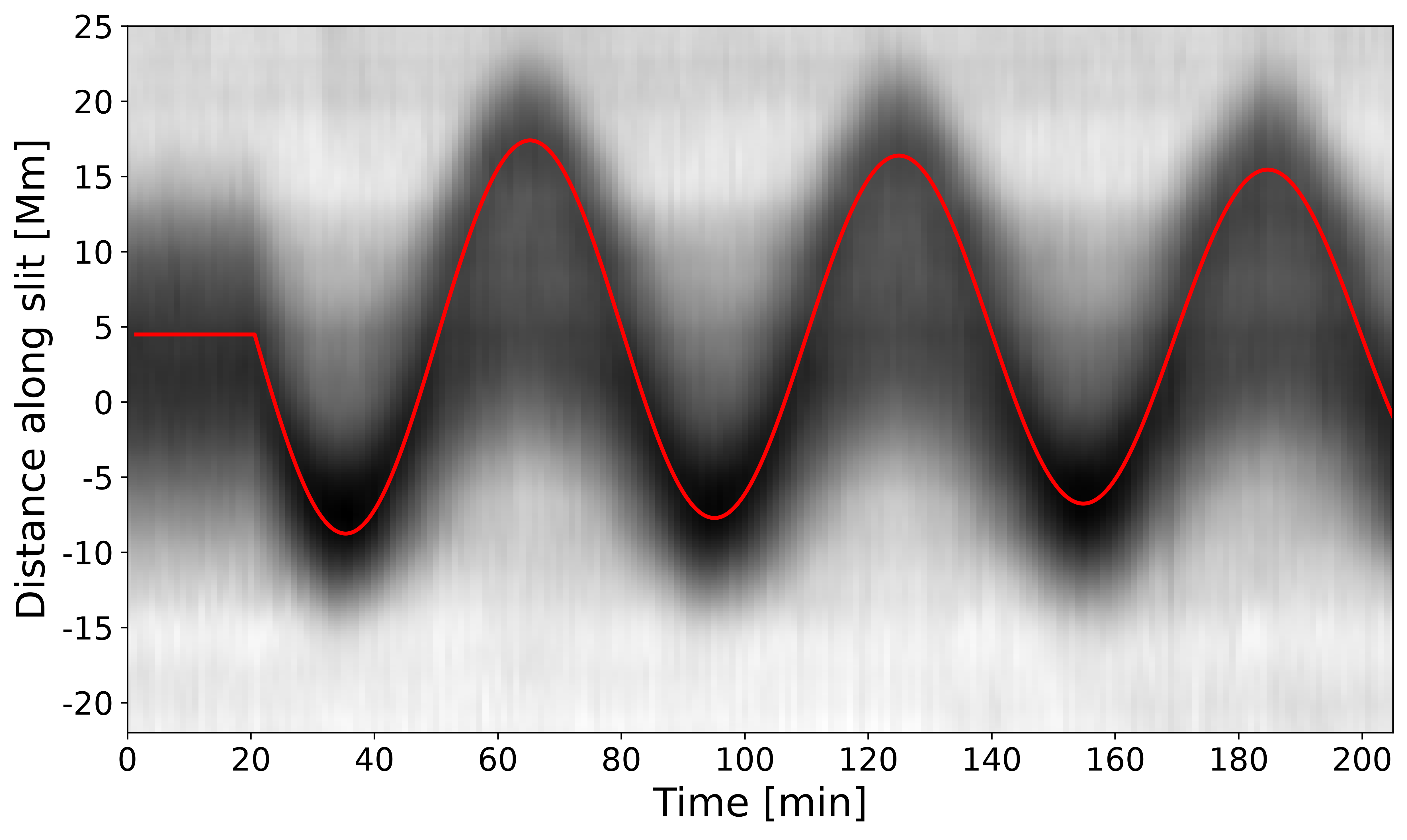}
    \caption{Time--distance diagram of the shared-period
    experiment. All filament threads oscillate with the
    same prescribed period, while the remaining simulation
    parameters are unchanged. The red curve represents
    the fitted damped-oscillation trajectory.}
    \label{fig:shared_period}
\end{figure}

Figure~\ref{fig:shared_period} shows the time--distance diagram obtained from this experiment, together with the fitted damped-oscillation trajectory. The slit used for this time-distance diagram is the exact same one from ~\ref{subsec:longitudinal}. The oscillatory displacement remains coherent throughout the sequence, with a more gradual decrease in amplitude than in the reference simulation. The fitted parameters are compared with those of the original experiment in Table~\ref{tab:dephasing_comparison}.

\begin{table}[!ht]
\caption{Comparison of the collective oscillation parameters
recovered from the reference simulation with distributed
thread periods and the experiment with a common period.
The uncertainties are formal one-standard-deviation
errors obtained from the fitting covariance matrices.}
\label{tab:dephasing_comparison}
\centering
\small
\begin{tabular}{@{}lcc@{}}
\hline\hline
Parameter & Distributed periods & Shared period \\
\hline
Amplitude (Mm) & $12.08\pm0.86$ & $13.62\pm0.35$ \\
Period (min) & $57.71\pm0.35$ & $59.74\pm0.11$ \\
Damping time (min) & $219\pm48$ & $732\pm132$ \\
$\tau/P$ & 3.80 & 12.26 \\
\hline
\end{tabular}
\end{table}

The recovered damping time increases from $219\pm48$~min in the reference simulation to $732\pm132$~min in the shared-period experiment, a factor of approximately 3.3. The latter value is also closer to the prescribed thread damping time of 906~min. This result supports the interpretation that the distribution of thread periods produces additional apparent damping of the collective H$\alpha$ displacement through progressive dephasing.

The shared-period experiment retains the effects of projection, image formation, and trajectory tracking, which can also influence the recovered damping time. Furthermore, the fitted damping time is considerably longer than the duration of the analyzed sequence, limiting the precision with which it can be determined. Nevertheless, the substantial increase in the recovered damping time when the period dispersion is removed demonstrates its important contribution to the decay of the
collective oscillatory signal.

The associated movies, of the \ha intensity sequence and derived velocity fields, for the shared period experiment are available online.

\end{appendix}
\end{document}